\documentclass[referee]{raa} 
\usepackage{lscape}           % referee version: for submission
\usepackage{natbib}
\bibstyle{model2-names.bst}
\usepackage{graphicx,times}             %for PS/EPS graphics inclusion, new
\usepackage{amssymb,amsmath}
\bibpunct{(}{)}{;}{a}{}{,}

\usepackage[a4paper=true,dvipdfm=true,pagebackref=true]{hyperref}
\hypersetup{colorlinks = true, linkcolor = green, anchorcolor = red, citecolor = blue, filecolor = red, pagecolor = red, urlcolor = red}

\begin{document}

   \title{Investigation of  the Relation between Space-Weather Parameters and Forbush Decreases Automatically Selected from Moscow and Apatity Cosmic Ray Stations during Solar Cycle 23. 
%\,$^*$
%\footnotetext{$*$ Supported by the National Natural Science Foundation of China.}
}
%   \subtitle{I. Place Your Subtitle Here}

   \volnopage{Vol.0 (20xx) No.0, 000--000}      %%preserved for Editor. DOn't remove!
   \setcounter{page}{1}          %%starting page, preserved for Editor. DOn't remove!

   \author{J. A. Alhassan
      \inst{1*}
   \and O. Okike%\'{\i}guez
      \inst{2}
   \and{ A. E. Chukwude}
      \inst{1}
   }
%% Here is an example of three authors come from different institutes.
%% For single author or all the authors from an institute, use "\inst{}" only

   \institute{Department of Physics and Astronomy, University of Nigeria, Nsukka, Nigeria \\
 %and {\it augustine.chukwude@unn.edu.ng}\\
%% Please give the E-mail address of the author, to whom future correspondence and
%% offprint requests will be sent.
        \and
             Department of Industrial Physics, Ebonyi State University, Abakaliki, Nigeria\\
{\it *Corr. Author:jibrin.alhassan@unn.edu.ng}\\ %{\it giftedlife2014@gmail.com}\\
%\corres{jibrin.alhassan@unn.edu.ng}
        %\and
            %Department of Physics and Astronomy, University of Nigeria, Nsukka, Nigeria {\it augustine.chukwude@unn.edu.ng}\\
\vs\no
   {\small Received~~20xx month day; accepted~~20xx~~month day}}

\abstract{ We present the results of an investigation of the relation between space-weather parameters and  cosmic ray (CR) intensity modulation using algorithm-selected Forbush decreases (FDs) from Moscow (MOSC) and Apatity (APTY) neutron monitor (NM) stations  during solar cycle 23. Our FD location program detected 408 and 383 FDs from MOSC and APTY NM stations respectively. A coincident  computer code employed in this work, detected 229 FDs that were observed at the same universal Time (UT) at the two stations. Out of the 229 simultaneous FDs, we formed a subset of 139 large  FDs(\%) $\leq-4$ at Moscow station. We peformed a  two dimensional regression analysis  between the FD magnitudes and the space-weather data on the two samples. We find that  there were significant
space-weather disturbances at the time of the CR flux depressions. The correlation between the space-weather parameters and  galactic cosmic ray (GCR) intensity decreases at the two NM stations are statistically significant. The implications of the present space-weather data on cosmic ray (CR) intensity depressions are highlighted. 
%The observed impact of solar-geophysical indices on FD magnitude does not appear to  depend on the  amplitude of cosmic ray (CR) intensity decreases.  %This  does not appear to depend on the amplitude of cosmic ray intensity decreases. 
\keywords{methods: data analysis - 
methods: statistical- Sun: coronal mass ejections (CMEs) - (Sun:) solar - terrestrial relations - (Sun:) solar wind - (ISM:) cosmic rays}
}

   \authorrunning{Alhassan, Okike \& Chukwude } 
           %author_head in even pages
   \titlerunning{ Relation between Space-Weather Parameters and Forbush Decreases}  % title_head in odd pages

   \maketitle
%% The author head (on even pages) and the title head (on odd pages) will be
%% automatically extracted from \author{} and \title{}. Whenever the title is too long,
%% you will be asked to supply a shorter one by inserting either \authorrunning{} or
%% \titlerunning{} before \maketitle. Anyway, you can specify your own heads.
%%
%%
%% Note: In the following text body of your manuscript, please note several differences from
%%       other major journals:
%% (1) \subsection{Please Capitalize the First Letter of Each Notional Word in Subsection Title}
%% (2) Please Capitalize the First Letter of Each Notional Word in all tables' captions

%\linenumbers
%________________________________________________ sections below
%
\section{Introduction}           %% first-level sections will be auto-capitalized
\label{sect:intro}

Forbush decreases (FDs)  are non-periodic short-term (hours to days) variabilities in the observed intensity  of galactic cosmic ray  (GCR) flux. This phenomenon that is  sometimes  regarded  as  rapid  and transient depression of 2\%\textendash 30\% of galactic cosmic ray intensity \citep{fo:1958,be:08,sank:2011}, was first reported by an American Physicist, Scott E. Forbush  over 80 years ago \citep{fo:38}. FDs caused by corotating high-speed solar wind streams (CSWSs)  from coronal holes (CHs) which corotate  with the sun are recurrent FDs, while events initiated by coronal mass ejections (CMEs) and their  interplanetary medium  signature (ICMEs) are regarded as non-recurrent or sporadic FDs \citep{ca:93,richard:2004, richard:2011, be:2014, richardson:2018}.  Recurrent FDs typically are symmetric and  have small magnitudes while large   magnitudes and asymmetric depressions characterizes non-recurrent FD group \citep{ Lockwood:1971,avbelov:2009,Melkumyan:2019}. Generally, solar wind disturbances in the form of  CMEs, ICMEs, corotating interaction regions (CIRs) and CSWSs structures determine cosmic ray intensity depressions on Earth \citep{ca:00,Dumbovic:2012, oki:20}. 

A large number of research works has intensively discussed FD phenomenon since its first observation, yet the origin, nature or connection of the
phenomenon with space-weather structures is still a subject of interest  among CR Scientists \citep{be:08, oki:20}. 
Association between FD amplitude, intensity of the interplanetary 
magnetic field (IMF),  solar wind speed (SWS) and geomagnetic storm time index (Dst) have been reported \citep[e.g.][]{burlaga:1984, ca:93, be:08, bel:2001, lin:2016,  oki:20, alh:2021} (\citet{alh:2021} will  be referred to as Paper I,  hereafter). \citet{bel:2001}, noted  that FDs and geomagnetic storm  are both traced to the disturbances in the interplanetary medium. By extension, disturbances in the solar wind, magnetosphere and FDs are triggerd by  the same  processes in the Sun. The study of these disturbances  are important in our understanding of the dynamics of solar-terrestrial environment since such activities can potentially result in  life threatening issues ranging from  satallite damage, communication failure to navigational difficulties \citep{jose:1981,balv:2011}. This underscores  the rationale for examining the impact of solar-geomagnetic characteristics on GCR intensity reductions.  

While SWS, IMF, Dst and other solar activity data could be sourced from some common  websites
(e.g., https:
//omniweb.gsfc.nasa.gov/html/ow data.html, http://wdc.kugi.kyoto-u.ac.jp/, http://cdaw.gsfc.nasa.
gov, ftp.ngdc.noaa.gov), FDs have to be calculated or sourced from http://spaceweather.izmiran.ru/eng/fds2005.htm. It is only the IZMIRAN group that have comprehensive FD catalogue obtained with  global survey method (GSM). The GSM  involves  computer codes and manual methods. The technique is better described as  semi-automated since a lot of manual work is incorporated \citep{be:2018}. Apart from the IZMIRAN group, other investigators have employed a few number of FDs either from literature \citep[e.g.][]{kris:08, lag:2009, laken:2011} or from manual technique \citep[e.g.][]{to:2001, ka:10, dra:2011}. 
The manual method is characteristically faced with a lot of demerits such as  time consuming nature of the process, difficulties in identification of low-amplitude FD events, inability to detect large number of FDs and absence of time domain in the model equation used to calculate FD event amplitude \citep[e.g.][]{ok:2020}. FD, being  the most spectacular variability in the GCR flux, \citet{rami:2013} suggested that  its  selection would require sophisticated technique. Computer software technique can  address the inherent limitations associated with the manual FD selection method. 

Emerging efforts to fully automate FD selection is already appearing in the literature \citep{ok:2019,ok:2020,okike:2020,oki:20,ogbos:2020, Light:2020}. With some technical improvements, we extend the  code employed by \citet{ok:2019} in our attempt to investigate the relation between solar-weather parameters and program selected FDs during solar cycle 23. FDs within this period is unique for study in several ways. Solar cycle 23 falls within the satellite period. It is well investigated  and   marked with high solar activity \citep{ok:2021}. Observations during this period are reported to be related to several coronal holes that  trigger high speed wind streams that presented many energetic events, especially during its declining phase \citep[e.g.][]{jean:2013,lin:2016}. 

\section{Data}
MOSC and APTY CR neutron monitor stations are among the old  observatories that have track record of continuous observations. MOSC has online data since 1958 to date while APTY has data since 1961 till date.   MOSC NM station is characterized by geographic location of  55.47$^{\circ}$N, 37.32$^{\circ}$E, geomagnetic cutoff rigidity of 2.43 GV  and altitude of 200 m with 24NM64 detector while APTY is located at 67.50$^{\circ}$N, 33.30$^{\circ}$E, rigidity of 0.57 GV and altitude of 177 m with 18NM64 detector.
CR pressure-corrected daily averaged data detected at  MOSC and APTY neutron monitor stations are downloaded from IZMIRAN  common website : \url{http://
cr0.izmiran.ru/common}.  
IMF, SWS and  Dst index data are obtained  from  https:
//omniweb.gsfc.nasa.gov/html/ow data.html.

\section{Algorithm FD Selection Method}
\subsection {FD Location  Progam}
In this section, we attempt  to  address some of the weakneses associated with the manual FD selection technique highlighted in section 1. The computer code  briefly described in this work, is an extension of  \citet{ok:2019} software. The technique in \citet{ok:2019} consists of modules for Fourier transformation  where the machine searches for FDs from the  transformed CR data. The present algorithm, written in  R program for statistical computing, a non-commercial software \citep{R:2014}, accepts CR raw data as its input signal instead of the high frequency input signal in \citet{ok:2019}. 
 Similar equation implemented by several authors \citep[e.g.][]{ har:2010} that manually calculate FD magnitude are used  in the present code for data normalization. 
 
The software  is equipped  to select FDs from both raw  and normalized CR  data. It is   programmed to search for depressions/turning points as well as  the time of the depressions in the  raw CR count. The depressions are signatures of FDs. Two subroutines  in the software are used to  calculate the FD magnitude  and the time of occurrences simultaneously. The baseline adopted determines the number of FDs that may be selected in a given period \citep[see][]{ok:2019}. Using a very small  baseline  ($\mathrm{CR} (\%)\leq -0.01$), as attempted by \citet{ok:2020}, \citet{ok:2021} and Paper I, the software identified  408  and 383 FDs respectively from MOSC and APTY neutron monitor stations. 

The IZMIRAN group  selected a total of 1346 Forbush Effects (FEs) 10 GV cosmic rays with GSM during this  period in which they assimilated CR counts from over fifty NM stations spread across the globe. The FDs detected at MOSC and APTY translates to  30\% and 29\% respectively of the number selected by the IZMIRAN group. The amplitude of the smallest FE is -0.2\%  whereas the
size of the least FD identified by the current code is -0.01\%. The  fully automated FD technique employed here has demonstrated   to be more efficient than semi-automated approach. Our software tracks FD minimum depression time while  GSM technique is based on  the event onset time. Using a large threshold defined as ($\mathrm{CR} (\%)\leq -4$), the number of FE  is 75 while FD is 139. This shows  that the large number differences between the two FD catalogues should lie majorly in the  small FD population. The observed  differences in number of FDs selected with both techniques could be due to  a number of factors that may include: altitude, rigidity, atmospheric depth, pressure,
temperature, relative humidity, local wind speed, the rotation of
the Earth with respect to the acceptance cone of the detectors,
latitudinal effects, instrumental variations, a station's sensitivity
to CR modulation, equatorial anisotropy, North-South anisotropy,
geomagnetic variations, snow, limited cone of acceptance, spurious
modulation, magnetospheric effects, differences in data resolution etc \citep[see][]{be:2018, oki:20}. 

FD  magnitude and time of occurrences selected with the location code are presented in Table 1.  In the Table, "S/N"  stands for serial number, "Date" for   FD dates at MOSC and APTY stations,  $FD_{MOSC}$(\%) for  FD  at MOSC and $FD_{APTY}$(\%) for FD  at APTY. 
 
\subsection{Simultaneous FD list} 
Simultaneity of  Forbush decreases has  been conceptualized variously by different investigators.  FD event is said to be simultaneous when the main phase  profile overlaps  in UT  \citep[e.g.][]{oh:08,lee:2013}. The complete FD event structure (onset, main phase, point of minimum reduction and the recovery phase) has been employed by \citet{ok:2011} in their attempt   to define FD event simultaneity.  In this submission, we define simultaneity  FDs  with respect to the event time of  minimum. The time of minimum of simultaneous FD would be detected at the same date at MOSC and APTY NMs. 

To  select FDs that are  coincident at MOSC and APTY  NM stations, we used a  simple  coincident computer code developed by \citet{oki:20}. FD coincident algorithm is a simple program written to select simultaneous  as well as non-simultaneous  FDs at two or more NM stations. Event time or magnitude may be used as the key search terms.  
These coincident FD events were selected  with respect to the magnitude of FDs at APTY station. The input data to the coincident algorithm are all the
FDs at each of the stations (408 at MOSC and 383 at   APTY, see Table 1). The output   show that  229 FDs  occured at the same  UT at the two stations. Out of  all the 229 simultaneous FDs selected by the coincident program,  we formed a subset of  139 large events at MOSC defined as FD(\%) $\leq-4$ \citep[see][]{okike:2020}.   %Thresholds commonly employed for the  the identification of large FDs FDs are CR( per cent) ≤ −3 (Pudovkin & Veretenenko 1995),CR( per cent) ≤ −3.5 (Oh et al. 2008) and CR( per cent) ≤ −5(Kristjansson et al. 2008(%) ≤ −4 at Moscow station
These datasets are  presented in Tables 2 and 3 alongside with their associated solar-geophysical parameters. The columns are organized as follows: "S/N", represents  serial number, "Date" stand for date of CR intensity depression, $FD_{MOSC}$(\%) and $FD_{APTY}$(\%) represent  MOSC and APTY  FDs respectively, "IMF" for interplanetary magnetic field, "SWS" for solar wind speed and "Dst" for disturbance storm time index. 

\section{ Results and Discussion}

%\subsection{Datasets Analyzed}
The FD catalogues  analysed in this work are all simultaneous FDs from MOSC and APTY CR stations (229 FDs), and large FD(\%) $\leq-4$ (139 FDs) at MOSC with their associated solar wind data, IMF and Dst.  The regression and correlation results are presented in Tables 4 and 5. The columns of Tables 4 and 5 are organized thus: S/N  for serial number, Parameters for each of the two continuous variables, $R^2$ indicate coefficient of determination, r is the Pearson's product moment correlation coefficient   and p-value is chance probability. The analysis of small amplitude FD(\%) $>$ -4 will be taken up in future investigation.
%  and large  FDs(\%) $\leq-4$ (139 FDs) at MOSC . The  datasets are presented in Tables 2 and 3. 

For the two FD catalogues (N = 229, N = 139), we found a connection between FD versus IMF. The $R^2$, r, and associated p-values for all the simultaneous events at MOSC (Fig. 1\textbf{a}), APTY (Fig.1\textbf{b}) and strong FDs at MOSC (Fig. 2\textbf{a})  are 0.13, 0.12 and 0.14, -0.36, -0.35 and -0.37,  $3.16\times 10^{-08}$, $8.53\times 10^{-08}$ and  $5.37\times 10^{-06}$ respectively. They all show a high level of significance at 95\% confidence level. The result show further that 13\%, 12\% and 14\% of CR intensity variations  observed in the two catalogues could be linked  to the effect of IMF intensity. While it may be inferred from Table \ref{table 2} that the magnitudes of the simultaneous FDs vary appreciably between the two stations, the close variation rate with respect to IMF observed (see Table \ref{table 4}) at MOSC and APTY CR stations seems to suggest that the simultaneity of FDs may not be dependent on the point of observation. Nevertheless, several other parameters such as  CME speed or transit speed, magnetospheric effects, sunspot number, current sheet tilt angle, solar magnetic turbulence level, CR anisotropy, heliospheric magnetic sector, rigidity, differences in the local time of the two stations, instrumental variations  and so on, which affect CR time-intensity  variations, should also be tested before reaching any definitive conclusions \citep{Cliver:1996,Singh:1997,Smith:1990,Wibberenz:2001,ow:2014, ok:2019a, okike:2020,oki:20}. In some past articles that analyzed the relationship between FDs and these physical parameters, the magnitude and timing of Forbush events were manually estimated. The present large event catalogue is an indication that only FD subsamples \citep[see][for a detailed discussion on the bias implications of a small sample of FD events]{lak:2012} were employed in some of the previous submissions that manually calculated FD event magnitude and timing \citep[e.g.][]{ba:75, richard:2004,ka:10}. These automated event catalogues provide opportunity for re-assessments as well as statistical analyses of the previously reported relations. Some of these investigations may be the focus of a future publication. 

One of the major pitfalls of the current algorithm is its inability to account for the influence of CR anisotropy on the timing and magnitude of FDs. Using a combination of fast Fourier transform and an FD location code, \citet{okike:2020} and \citet{okike:2020c} provided, for the first time, an empirical evidence of the significant differences that might exist between FDs identified from unprocessed and those selected from Fourier transformed CR data. Before making a firm statement on the similar correlation between FDs and IMF at APTY and MOSC stations as well as other relations reported here, the present analysis should be repeated using FD data calculated from Fourier transformed CR flux variation.

%In Figure 2 panel \textbf{a}, we display the scatter plot of  the magnitude of $FD_{MOSC}$ versus the IMF for the large amplitude simultaneous FDs at MOSC data. The analysis yield  $R^2$ $\sim {0.14}$,  $r\sim {-0.37}$ with p-value of $5.37\times 10^{-06}$ significant at 95\% confidence level. From the $R^2$ value, We deduce that 14\% of the CR flux variation can be accounted for by interplanetary magnetic field intensity.

Based on strong FDs selected using a CR(\%) $\leq-3$ large baseline, \citet{ogbos:2020} found FD versus IMF correlation coefficient $r\sim {-0.34}$ statistically significant at 90\% significant level. The present regression analysis result   compare favourably well with their finding. This is consistent with the fact  that enhanched IMF hinders CR intensity propagation to the Earth and thus results in high magnitude FD detection \citep{lin:2016}.

%The statistical analyses results of the two FD catalogues reveal that on average, simultaneous FDs at the two NM stations and large amplitude simultaneous FDs at MOSC show comparable  tendencies  of linear relationship with  solar wind data and geomagnetic storm activity parameters as presented in Tables 4 and 5.

%of CR intensity variation at MOSC and APTY stations respectively are due to IMF intensity.

The plots of $FD_{MOSC}$-$SWS$, $FD_{APTY}$-$SWS$ for all simultaneous FD list and $FD_{MOSC}$-$SWS$ for large FDs are shown in Fig.1\textbf{c}, Fig.1\textbf{d} and Fig.2\textbf{b} respectively. Note that  their respective $R^2$, r and p-values are 0.17, 0.11 and 0.23, 0.41, 0.34 and 0.48, $1.27\times 10^{-10}$, $1.45\times 10^{-07}$ and $2.13\times 10^{-09}$. The p-values indicate that the results are  statistically significant at 95\% confidence level. The values of $R^2$ suggest that  variation (17\%), (11\%) and (23\%) in SWS play some roles that cannot be ignored in comparison with other solar agents that might influence large amplitude FDs. 

%For the large FD dataset, we show the  graph of $FD_{MOSC}$-$SWS$  in panel \textbf{b} of Figure 2. The regression analysis gives $R^2$ $\sim {0.23}$,  $r\sim {-0.48}$ with p-value of $2.13\times 10^{-09}$ statistically significant at 95\% confidence level. The value of  $R^2$ suggest that reasonable variation (23\%) in FDs could be associated with the effect of SWS. This result could also mean that SWS play reasonable role in comparison with other solar agents that might influence large amplitude FDs.

In an earlier work, \citet{iucci:79} using averaged hourly neutron monitor data from Alert, Deep River, Goose Bay and Inuvik from 1964-1974, found that FD amplitude are not directly related to SWS. Our results in the two datasets is contrary to  this submission. We found evidence of FD versus SWS relation significant at 95\% confidence level. This disparity  could be a pointer to the merits of fully automated FD data over manual approach employed by \citet{iucci:79}. In separate works, \citet{be:2014}, \citet{bad:2015}, \citet{lin:2016} and \citet{ogbos:2020} reported FD-SWS correlation coefficient of 0.65, 0.58, 0.40 and 0.81 respectively. Our result  for the two datasets on average tend to reflect the submissions of \citet{bad:2015} and  \citet{lin:2016} but lower than that of \citet{be:2014} and \citet{ogbos:2020}. The present result which is significant at 95\% confidence level indicate a strong impact of solar wind structure on GCR flux modulations. 

$FD_{MOSC}$-$Dst$, $FD_{APTY}$-$Dst$ for all simultaneous FD data and $FD_{MOSC}$-$Dst$ for strong FD data are plotted as shown in (Fig.1\textbf{e}), (Fig.1\textbf{f}) and (Fig.2\textbf{c}) respectively. The analysis yield their corresponding $R^2$, r and p-values as 0.16, 0.14 and 0.13, 0.40, 0.37 and 0.36,  $6.51\times 10^{-10}$, $6.20\times 10^{-09}$ and $1.34\times 10^{-05}$. These results indicate that 16\%, 14\% and 13\% of CR flux modulation at MOSC and APTY neutron monitor stations can be attributed to disturbance storm time index. The chance probability value further show that the results are highly significant at 95\% confidence level. 

%In panel \textbf{c} of Figure 2, we display the plot of $FD_{MOSC}$-$Dst$ for the large FD data. From the correlation and regression analyses, we obtained $R^2$ $\sim {0.13}$,  $r\sim {-0.36}$ and  p-value of $1.34\times 10^{-05}$ significant at 95\% confidence level. The result of  $R^2$ indicates that 13\% of the GCR intensity modulation is attributed to geomagnetic storm index.

\citet{bel:2001}, perfomed a regression analysis for 1428 events and obtained a correlation coefficient of ${<}$ 0.42 between the FD amplitude and the Dst-index.  The results we find for FD versus Dst parameters for the two datasets is in close agreement  with their finding. Recently, Paper I reported  a correlation coefficient of 0.46 for FD-Dst relation based on 129 large CR(\%) $\leq-3$ data from Oulu NM station. While we note that  the result obtained from the current data is in close agreement with the result of Paper I, we remark that  with respect to Dst, the properties of CR flux modulation appears to be   dependent on the point of observation.

%\subsection{Correlation results of large amplitude simultaneous FDs ((\%) $\leq-4)$ at MOSC}In Table 3, we present the large magnitude FDs at MOSC and their corresponding solar/geophysical parameters. The regression and correlation analyses results  are shown in Table 5. In Figure 2 panel \textbf{a}, we display the scatter plot of  the magnitude of $FD_{MOSC}$ versus the IMF. The analysis yield  $R^2$ $\sim {0.14}$,  $r\sim {-0.37}$ with p-value of $5.37\times 10^{-06}$ significant at 95\% confidence level. From the $R^2$ value, We deduce that 14\% of the CR flux variation can be accounted for by interplanetary magnetic field intensity.

%The  graph of $FD_{MOSC}$-$SWS$ is shown in panel \textbf{b} of Figure 2. The regression analysis gives $R^2$ $\sim {0.23}$,  $r\sim {-0.48}$ with p-value of $2.13\times 10^{-09}$ statistically significant at 95\% confidence level. The value of  $R^2$ suggest that reasonable variation (23\%) in FDs could be associated with the effect of SWS. This result could also mean that SWS play reasonable role in comparison with other solar agents that might influence large amplitude FDs. 
% of CR intensity variation is contributed by  solar wind.

%In panel \textbf{c} of Figure 2, we display the plot of $FD_{MOSC}$-$Dst$. From the correlation and regression analyses, we obtained $R^2$ $\sim {0.13}$,  $r\sim {-0.36}$ and  p-value of $1.34\times 10^{-05}$ significant at 95\% confidence level. The result of  $R^2$ indicates that 13\% of the GCR intensity modulation is attributed to geomagnetic storm index.
\section{Summary and Conclusions} 
Investigation of solar-weather parameters responsible for GCR variations on Earth remain a subject of interest. Several works on this subject employ correlation and regression statistical approach but often fail to carry out significance test that accounts for autocorrelation in solar-geophysical data \citep{belov:2001, lin:2016}. The selection of FDs being a key parameter is of primary importance in the study of solar-terrestrial connection. Many investigators employ either manual or semi-automated methods in identifying FDs. In this work, we have deployed a fully FD location algorithm to both calculate the amplitude of FDs as well as detecting the number of FDs that occurred between 1996-2005. The FD location code employed here is capable of selecting FD magnitude of -0.01\%. The correlation and regression results show that there were significant
space-weather perturbations at the time of the CR flux depressions. All the relations are statistically highly significant.   The two dimensional analysis carried out in this work reveal that  that there are  other factors other than the IMF, SWS and Dst that
control the amount of CR flux that arrives on Earth.

The observed high statistical significance of the  correlation between FDs, solar wind data and geophysical parameters could imply  that SWS, IMF intensity and Dst have the same causative agent. Their impact on GCR intensity variations are significant. Program selected FDs has shown that  solar activity parameters and geomagnetic storm characteristics investigated in this work, are  important factors that drive  variations of FDs  \citep{richard:2004, richard:2011}. 

\begin{table}[ht]

\caption{ \textbf{ Selected FDs at MOSC  and APTY Stations from 1996-2005}}
\label{table 1}
\centering
\begin{tabular}{rlrlr}
  \hline
S/N & Date & $FD_{MOSC}$(\%) & Date & $FD_{APTY}$(\%) \\ 
  \hline
1 & 1998-05-02 & -1.78 & 1998-05-02 & -1.34 \\ 
  2 & 1998-08-27 & -3.16 & 1998-05-04 & -1.57 \\ 
  3 & 1998-09-25 & -2.16 & 1998-08-27 & -2.08 \\ 
  4 & 1999-01-24 & -2.70 & 1999-02-18 & -2.96 \\ 
  5 & 1999-02-18 & -3.46 & 1999-08-22 & -1.61 \\ 
  6 & 1999-06-27 & -0.52 & 1999-08-25 & -0.53 \\ 
  7 & 1999-08-20 & -2.04 & 1999-09-13 & -0.27 \\ 
  8 & 1999-08-22 & -2.57 & 1999-09-16 & -1.05 \\ 
  9 & 1999-08-25 & -1.72 & 1999-09-20 & -0.68 \\ 
  10 & 1999-09-05 & -0.89 & 1999-09-25 & -0.01 \\ 
  11 & 1999-09-07 & -0.64 & 1999-09-29 & -1.50 \\ 
  12 & 1999-09-09 & -0.95 & 1999-10-12 & -1.57 \\ 
  13 & 1999-09-16 & -1.76 & 1999-10-15 & -3.16 \\ 
  14 & 1999-09-18 & -1.68 & 1999-10-17 & -3.34 \\ 
  15 & 1999-09-21 & -1.86 & 1999-10-22 & -3.26 \\ 
  16 & 1999-09-25 & -0.78 & 1999-10-24 & -3.31 \\ 
  17 & 1999-09-29 & -1.53 & 1999-10-29 & -2.38 \\ 
  18 & 1999-10-03 & -1.45 & 1999-11-01 & -2.86 \\ 
  19 & 1999-10-05 & -1.33 & 1999-11-09 & -2.37 \\ 
  20 & 1999-10-12 & -2.20 & 1999-11-18 & -4.27 \\ 
  21 & 1999-10-16 & -3.84 & 1999-11-20 & -4.83 \\ 
  22 & 1999-10-21 & -3.32 & 1999-11-27 & -2.93 \\ 
  23 & 1999-10-23 & -3.19 & 1999-11-29 & -2.20 \\ 
  24 & 1999-10-25 & -3.23 & 1999-12-02 & -3.64 \\ 
  25 & 1999-11-01 & -2.93 & 1999-12-13 & -7.49 \\ 
  26 & 1999-11-06 & -2.19 & 1999-12-27 & -4.53 \\ 
  27 & 1999-11-09 & -2.43 & 2000-01-04 & -3.89 \\ 
  28 & 1999-11-14 & -3.65 & 2000-01-07 & -4.64 \\ 
  29 & 1999-11-18 & -3.72 & 2000-01-13 & -2.94 \\ 
  30 & 1999-11-20 & -3.63 & 2000-01-16 & -2.45 \\ 
  31 & 1999-11-22 & -3.79 & 2000-01-21 & -2.24 \\ 
  32 & 1999-12-02 & -3.29 & 2000-01-26 & -3.61 \\ 
  33 & 1999-12-07 & -1.78 & 2000-01-28 & -3.34 \\ 
  34 & 1999-12-13 & -7.47 & 2000-01-31 & -4.05 \\ 
  35 & 1999-12-27 & -4.35 & 2000-02-07 & -4.44 \\ 
  36 & 1999-12-31 & -2.81 & 2000-02-12 & -6.44 \\ 
  37 & 2000-01-05 & -1.82 & 2000-02-21 & -4.67 \\ 
  38 & 2000-01-07 & -1.73 & 2000-03-02 & -6.97 \\ 
  39 & 2000-01-09 & -1.67 & 2000-03-08 & -5.01 \\ 
  40 & 2000-01-13 & -1.68 & 2000-03-13 & -5.07 \\ 
  41 & 2000-01-24 & -1.28 & 2000-03-20 & -4.93 \\ 
  42 & 2000-01-29 & -1.28 & 2000-03-24 & -6.37 \\ 
  43 & 2000-01-31 & -1.22 & 2000-03-30 & -5.79 \\ 
  44 & 2000-02-02 & -0.81 & 2000-04-07 & -6.84 \\ 
  45 & 2000-02-09 & -1.94 & 2000-04-14 & -3.44 \\ 
  46 & 2000-02-12 & -5.31 & 2000-04-17 & -4.20 \\ 
  47 & 2000-02-17 & -3.28 & 2000-04-20 & -4.60 \\ 
  48 & 2000-02-19 & -3.04 & 2000-04-22 & -4.64 \\
\hline
\end{tabular}
\end{table}

\begin{table}[ht]
%\caption{ \textbf{ FD List from MOSC FD1(\%) and APTY FD2(\%) during 1996-2005}}
\centering
\label*{table 1}
\begin{tabular}{rlrlr}
  \hline
S/N & Date & $FD_{MOSC}$(\%) & Date & $FD_{APTY}$(\%) \\ 
  \hline
 
  49 & 2000-02-21 & -3.90 & 2000-04-24 & -5.33 \\ 
  50 & 2000-03-01 & -4.19 & 2000-04-28 & -3.72 \\ 
  51 & 2000-03-09 & -3.62 & 2000-05-03 & -6.53 \\ 
  52 & 2000-03-15 & -3.65 & 2000-05-08 & -7.04 \\ 
  53 & 2000-03-20 & -3.65 & 2000-05-15 & -6.70 \\ 
  54 & 2000-03-24 & -5.18 & 2000-05-24 & -10.78 \\ 
  55 & 2000-03-30 & -4.57 & 2000-05-30 & -6.84 \\
  56 & 2000-04-04 & -4.65 & 2000-06-09 & -11.70 \\ 
  57 & 2000-04-07 & -6.15 & 2000-06-20 & -8.40 \\ 
  58 & 2000-04-14 & -2.96 & 2000-06-24 & -8.78 \\ 
  59 & 2000-04-17 & -3.90 & 2000-06-26 & -8.63 \\ 
  60 & 2000-04-20 & -4.70 & 2000-07-01 & -6.15 \\ 
  61 & 2000-04-24 & -4.53 & 2000-07-05 & -6.22 \\ 
  62 & 2000-05-03 & -4.35 & 2000-07-11 & -8.63 \\ 
  63 & 2000-05-08 & -5.12 & 2000-07-13 & -10.95 \\ 
  64 & 2000-05-15 & -4.31 & 2000-07-16 & -18.13 \\ 
  65 & 2000-05-24 & -8.87 & 2000-07-20 & -13.62 \\ 
  66 & 2000-05-30 & -5.66 & 2000-07-22 & -12.95 \\ 
  67 & 2000-06-09 & -10.62 & 2000-07-28 & -10.58 \\ 
  68 & 2000-06-20 & -6.92 & 2000-08-06 & -10.59 \\ 
  69 & 2000-06-24 & -7.21 & 2000-08-12 & -11.70 \\ 
  70 & 2000-06-26 & -7.25 & 2000-08-15 & -10.24 \\ 
  71 & 2000-07-02 & -4.86 & 2000-08-24 & -7.18 \\ 
  72 & 2000-07-05 & -4.81 & 2000-08-29 & -7.06 \\ 
  73 & 2000-07-11 & -7.44 & 2000-09-03 & -7.04 \\ 
  74 & 2000-07-16 & -16.72 & 2000-09-09 & -8.04 \\ 
  75 & 2000-07-20 & -12.51 & 2000-09-16 & -7.99 \\ 
  76 & 2000-07-29 & -9.48 & 2000-09-18 & -12.03 \\ 
  77 & 2000-08-06 & -8.90 & 2000-09-26 & -6.25 \\ 
  78 & 2000-08-12 & -9.76 & 2000-09-29 & -6.74 \\ 
  79 & 2000-08-25 & -5.23 & 2000-10-01 & -6.64 \\ 
  80 & 2000-08-29 & -5.37 & 2000-10-05 & -6.62 \\ 
  81 & 2000-09-03 & -5.64 & 2000-10-07 & -6.99 \\ 
  82 & 2000-09-07 & -5.96 & 2000-10-14 & -6.14 \\ 
  83 & 2000-09-09 & -6.30 & 2000-10-17 & -4.55 \\ 
  84 & 2000-09-15 & -6.01 & 2000-10-22 & -4.19 \\ 
  85 & 2000-09-18 & -9.62 & 2000-10-25 & -3.70 \\ 
  86 & 2000-09-25 & -4.18 & 2000-10-29 & -8.11 \\ 
  87 & 2000-09-29 & -4.34 & 2000-11-01 & -7.25 \\ 
  88 & 2000-10-01 & -4.01 & 2000-11-04 & -7.11 \\ 
  89 & 2000-10-05 & -5.01 & 2000-11-07 & -9.85 \\ 
  90 & 2000-10-07 & -5.41 & 2000-11-11 & -8.44 \\ 
  91 & 2000-10-14 & -4.54 & 2000-11-16 & -7.51 \\ 
  92 & 2000-10-20 & -2.93 & 2000-11-24 & -7.07 \\ 
  93 & 2000-10-29 & -7.10 & 2000-11-29 & -12.28 \\ 
  94 & 2000-11-01 & -6.60 & 2000-12-03 & -9.43 \\ 
  95 & 2000-11-03 & -6.19 & 2000-12-06 & -9.19 \\ 
  96 & 2000-11-07 & -8.68 & 2000-12-11 & -7.06 \\ 
  97 & 2000-11-11 & -7.02 & 2000-12-17 & -5.74 \\ 
\hline
\end{tabular}
\end{table}

\begin{table}[ht]
%\caption{ \textbf{ FD List from MOSC FD1(\%) and APTY FD2(\%) during 1996-2005}}
\centering
\label*{table 1}
\begin{tabular}{rlrlr}
  \hline
S/N & Date & $FD_{MOSC}$(\%) & Date & $FD_{APTY}$(\%) \\ 
  \hline

  98 & 2000-11-16 & -6.49 & 2000-12-19 & -6.04 \\ 
  99 & 2000-11-23 & -5.20 & 2000-12-23 & -6.74 \\ 
  100 & 2000-11-29 & -10.52 & 2000-12-25 & -6.97 \\ 
  101 & 2000-12-03 & -7.54 & 2000-12-27 & -6.95 \\ 
  102 & 2000-12-11 & -5.70 & 2000-12-30 & -5.85 \\ 
  103 & 2000-12-25 & -6.06 & 2001-01-03 & -6.34 \\ 
  104 & 2000-12-27 & -6.29 & 2001-01-05 & -6.12 \\ 
  105 & 2001-01-03 & -4.85 & 2001-01-09 & -6.40 \\ 
  106 & 2001-01-05 & -4.86 & 2001-01-13 & -5.73 \\ 
  107 & 2001-01-07 & -4.79 & 2001-01-19 & -5.53 \\ 
  108 & 2001-01-09 & -4.89 & 2001-01-25 & -6.74 \\ 
  109 & 2001-01-17 & -5.05 & 2001-01-31 & -5.12 \\ 
  110 & 2001-01-24 & -5.57 & 2001-02-09 & -3.94 \\ 
  111 & 2001-02-01 & -3.96 & 2001-02-11 & -3.63 \\ 
  112 & 2001-02-07 & -2.52 & 2001-02-14 & -5.15 \\ 
  113 & 2001-02-10 & -2.59 & 2001-02-20 & -3.71 \\ 
  114 & 2001-02-13 & -3.09 & 2001-02-25 & -2.56 \\ 
  115 & 2001-02-15 & -2.62 & 2001-02-28 & -2.37 \\ 
  116 & 2001-02-21 & -1.76 & 2001-03-04 & -3.83 \\ 
  117 & 2001-03-04 & -2.43 & 2001-03-13 & -1.04 \\ 
  118 & 2001-03-20 & -1.43 & 2001-03-20 & -3.00 \\ 
  119 & 2001-03-28 & -3.16 & 2001-03-24 & -1.56 \\ 
  120 & 2001-04-01 & -5.22 & 2001-03-28 & -4.60 \\ 
  121 & 2001-04-05 & -6.53 & 2001-04-01 & -6.96 \\ 
  122 & 2001-04-09 & -8.15 & 2001-04-05 & -7.04 \\ 
  123 & 2001-04-12 & -14.30 & 2001-04-09 & -8.75 \\ 
  124 & 2001-04-16 & -7.08 & 2001-04-12 & -14.88 \\ 
  125 & 2001-04-19 & -5.83 & 2001-04-16 & -7.78 \\ 
  126 & 2001-04-22 & -4.85 & 2001-04-19 & -6.97 \\ 
  127 & 2001-04-25 & -3.49 & 2001-04-22 & -5.44 \\ 
  128 & 2001-04-29 & -7.99 & 2001-04-26 & -4.45 \\ 
  129 & 2001-05-12 & -2.25 & 2001-04-29 & -9.52 \\ 
  130 & 2001-05-16 & -2.35 & 2001-05-09 & -4.03 \\ 
  131 & 2001-05-25 & -4.02 & 2001-05-12 & -4.18 \\ 
  132 & 2001-05-28 & -6.17 & 2001-05-15 & -4.20 \\ 
  133 & 2001-06-03 & -3.18 & 2001-05-25 & -5.73 \\ 
  134 & 2001-06-09 & -3.06 & 2001-05-28 & -8.11 \\ 
  135 & 2001-06-12 & -3.00 & 2001-06-03 & -4.37 \\ 
  136 & 2001-06-20 & -3.54 & 2001-06-09 & -4.30 \\ 
  137 & 2001-06-26 & -2.51 & 2001-06-12 & -4.05 \\ 
  138 & 2001-06-29 & -2.44 & 2001-06-20 & -4.46 \\ 
  139 & 2001-07-04 & -3.52 & 2001-06-26 & -3.90 \\ 
  140 & 2001-07-11 & -2.37 & 2001-06-29 & -4.12 \\ 
  141 & 2001-07-14 & -1.64 & 2001-07-05 & -5.22 \\ 
  142 & 2001-07-17 & -2.04 & 2001-07-09 & -4.15 \\ 
  143 & 2001-07-19 & -2.10 & 2001-07-17 & -2.83 \\ 
  144 & 2001-07-21 & -2.27 & 2001-07-20 & -3.26 \\ 
  145 & 2001-07-25 & -3.87 & 2001-07-26 & -4.04 \\ 
\hline
\end{tabular}
\end{table}

\begin{table}[ht]
%\caption{ \textbf{ FD List from MOSC FD1(\%) and APTY FD2(\%) during 1996-2005}}
\centering
\label*{table 1}
\begin{tabular}{rlrlr}
  \hline
S/N & Date & $FD_{MOSC}$(\%) & Date & $FD_{APTY}$(\%) \\
  \hline

  146 & 2001-07-30 & -3.12 & 2001-07-30 & -4.23 \\ 
  147 & 2001-08-03 & -4.22 & 2001-08-03 & -5.30 \\ 
  148 & 2001-08-06 & -4.24 & 2001-08-06 & -5.22 \\ 
  149 & 2001-08-10 & -3.00 & 2001-08-10 & -3.94 \\ 
  150 & 2001-08-14 & -3.11 & 2001-08-14 & -4.08 \\ 
  151 & 2001-08-19 & -4.79 & 2001-08-18 & -6.62 \\ 
  152 & 2001-08-23 & -4.45 & 2001-08-24 & -5.92 \\ 
  153 & 2001-08-29 & -8.46 & 2001-08-29 & -9.67 \\ 
  154 & 2001-09-07 & -3.79 & 2001-09-07 & -5.53 \\ 
  155 & 2001-09-14 & -3.25 & 2001-09-14 & -4.29 \\ 
  156 & 2001-09-19 & -3.01 & 2001-09-16 & -4.57 \\ 
  157 & 2001-09-26 & -8.82 & 2001-09-19 & -4.48 \\ 
  158 & 2001-10-02 & -8.58 & 2001-09-26 & -10.48 \\ 
  159 & 2001-10-09 & -5.48 & 2001-10-02 & -10.14 \\ 
  160 & 2001-10-12 & -6.31 & 2001-10-09 & -5.96 \\ 
  161 & 2001-10-22 & -5.61 & 2001-10-12 & -7.45 \\ 
  162 & 2001-10-28 & -4.82 & 2001-10-22 & -8.44 \\ 
  163 & 2001-11-07 & -6.56 & 2001-10-28 & -7.01 \\ 
  164 & 2001-11-14 & -2.26 & 2001-11-03 & -3.74 \\ 
  165 & 2001-11-25 & -8.61 & 2001-11-07 & -9.15 \\ 
  166 & 2001-12-01 & -2.33 & 2001-11-13 & -4.34 \\ 
  167 & 2001-12-07 & -4.49 & 2001-11-15 & -4.30 \\ 
  168 & 2001-12-17 & -3.95 & 2001-11-22 & -5.42 \\ 
  169 & 2001-12-22 & -0.94 & 2001-11-25 & -10.76 \\ 
  170 & 2001-12-29 & -2.41 & 2001-12-04 & -5.82 \\ 
  171 & 2002-01-03 & -7.86 & 2001-12-07 & -5.92 \\ 
  172 & 2002-01-11 & -6.33 & 2001-12-17 & -5.79 \\ 
  173 & 2002-01-21 & -3.51 & 2001-12-22 & -3.49 \\ 
  174 & 2002-01-24 & -3.28 & 2001-12-27 & -3.15 \\ 
  175 & 2002-02-02 & -5.22 & 2001-12-29 & -4.25 \\ 
  176 & 2002-02-11 & -0.44 & 2002-01-03 & -9.25 \\ 
  177 & 2002-02-15 & -0.34 & 2002-01-11 & -8.44 \\ 
  178 & 2002-02-19 & -0.74 & 2002-01-21 & -5.23 \\ 
  179 & 2002-02-23 & -1.67 & 2002-01-29 & -6.88 \\ 
  180 & 2002-02-26 & -1.70 & 2002-02-01 & -6.49 \\ 
  181 & 2002-03-02 & -2.03 & 2002-02-05 & -6.07 \\ 
  182 & 2002-03-05 & -1.83 & 2002-02-12 & -3.55 \\ 
  183 & 2002-03-09 & -0.71 & 2002-02-16 & -3.46 \\ 
  184 & 2002-03-12 & -1.96 & 2002-02-19 & -3.82 \\ 
  185 & 2002-03-16 & -1.85 & 2002-02-24 & -4.63 \\ 
  186 & 2002-03-24 & -6.37 & 2002-03-01 & -5.03 \\ 
  187 & 2002-03-30 & -3.43 & 2002-03-05 & -4.01 \\ 
  188 & 2002-04-01 & -3.34 & 2002-03-12 & -3.61 \\ 
  189 & 2002-04-06 & -1.94 & 2002-03-16 & -3.61 \\ 
  190 & 2002-04-12 & -2.92 & 2002-03-20 & -7.80 \\ 
  191 & 2002-04-15 & -1.72 & 2002-03-25 & -8.45 \\ 
  192 & 2002-04-18 & -4.86 & 2002-03-28 & -6.40 \\ 
  193 & 2002-04-20 & -5.63 & 2002-03-30 & -5.94 \\
\hline
\end{tabular}
\end{table}

\begin{table}[ht]
%\caption{ \textbf{ FD List from MOSC FD1(\%) and APTY FD2(\%) during 1996-2005}}
\centering
\label*{table 1}
\begin{tabular}{rlrlr}
  \hline
S/N & Date & $FD_{MOSC}$(\%) & Date & $FD_{APTY}$(\%) \\ 
  \hline
 
  194 & 2002-04-24 & -4.95 & 2002-04-06 & -4.16 \\ 
  195 & 2002-04-30 & -3.13 & 2002-04-12 & -5.15 \\ 
  196 & 2002-05-08 & -1.64 & 2002-04-15 & -4.11 \\ 
  197 & 2002-05-12 & -3.10 & 2002-04-18 & -6.55 \\ 
  198 & 2002-05-15 & -3.73 & 2002-04-20 & -7.53 \\ 
  199 & 2002-05-23 & -5.83 & 2002-04-24 & -7.33 \\ 
  200 & 2002-05-28 & -4.53 & 2002-05-04 & -2.94 \\ 
  201 & 2002-06-03 & -2.92 & 2002-05-08 & -3.20 \\ 
  202 & 2002-06-08 & -2.90 & 2002-05-13 & -5.04 \\ 
  203 & 2002-06-11 & -3.84 & 2002-05-15 & -5.49 \\ 
  204 & 2002-06-16 & -2.95 & 2002-05-20 & -6.42 \\ 
  205 & 2002-06-19 & -3.65 & 2002-05-23 & -7.06 \\ 
  206 & 2002-06-25 & -2.42 & 2002-05-27 & -6.25 \\ 
  207 & 2002-06-29 & -1.62 & 2002-06-07 & -4.89 \\ 
  208 & 2002-07-02 & -2.22 & 2002-06-11 & -5.59 \\ 
  209 & 2002-07-08 & -3.65 & 2002-06-19 & -5.84 \\ 
  210 & 2002-07-11 & -3.80 & 2002-06-24 & -4.00 \\ 
  211 & 2002-07-20 & -6.52 & 2002-06-29 & -3.56 \\ 
  212 & 2002-07-23 & -5.46 & 2002-07-04 & -4.08 \\ 
  213 & 2002-07-30 & -9.02 & 2002-07-08 & -5.07 \\ 
  214 & 2002-08-02 & -9.65 & 2002-07-12 & -5.10 \\ 
  215 & 2002-08-07 & -6.29 & 2002-07-18 & -6.63 \\ 
  216 & 2002-08-09 & -5.63 & 2002-07-20 & -8.32 \\ 
  217 & 2002-08-20 & -6.95 & 2002-07-23 & -7.15 \\ 
  218 & 2002-08-23 & -6.80 & 2002-07-30 & -10.43 \\ 
  219 & 2002-08-28 & -7.57 & 2002-08-03 & -10.36 \\ 
  220 & 2002-09-03 & -5.58 & 2002-08-09 & -7.41 \\ 
  221 & 2002-09-08 & -6.06 & 2002-08-20 & -8.66 \\ 
  222 & 2002-09-12 & -5.11 & 2002-08-23 & -8.74 \\ 
  223 & 2002-09-24 & -5.87 & 2002-08-28 & -9.37 \\ 
  224 & 2002-09-28 & -4.53 & 2002-08-30 & -8.78 \\ 
  225 & 2002-10-01 & -4.86 & 2002-09-04 & -7.29 \\ 
  226 & 2002-10-03 & -4.91 & 2002-09-08 & -7.73 \\ 
  227 & 2002-10-11 & -1.90 & 2002-09-11 & -7.22 \\ 
  228 & 2002-10-13 & -2.60 & 2002-09-19 & -5.94 \\ 
  229 & 2002-10-21 & -6.40 & 2002-09-24 & -6.86 \\ 
  230 & 2002-10-25 & -5.27 & 2002-09-28 & -5.74 \\ 
  231 & 2002-10-31 & -3.72 & 2002-10-01 & -5.93 \\ 
  232 & 2002-11-03 & -5.26 & 2002-10-03 & -6.86 \\ 
  233 & 2002-11-05 & -6.09 & 2002-10-09 & -4.25 \\ 
  234 & 2002-11-12 & -7.79 & 2002-10-13 & -3.86 \\ 
  235 & 2002-11-18 & -9.28 & 2002-10-22 & -7.40 \\ 
  236 & 2002-11-25 & -4.59 & 2002-10-25 & -7.10 \\ 
  237 & 2002-11-27 & -5.47 & 2002-10-29 & -6.23 \\ 
  238 & 2002-12-08 & -4.74 & 2002-11-03 & -7.26 \\ 
  239 & 2002-12-14 & -5.22 & 2002-11-05 & -8.32 \\ 
  240 & 2002-12-16 & -5.34 & 2002-11-12 & -9.43 \\ 
\hline
\end{tabular}
\end{table}

\begin{table}[ht]
%\caption{ \textbf{ FD List from MOSC FD1(\%) and APTY FD2(\%) during 1996-2005}}
\centering
\label*{table 1}
\begin{tabular}{rlrlr}
  \hline
S/N & Date & $FD_{MOSC}$(\%) & Date & $FD_{APTY}$(\%) \\ 
  \hline

  241 & 2002-12-18 & -5.30 & 2002-11-18 & -10.98 \\ 
  242 & 2002-12-20 & -5.87 & 2002-11-25 & -5.88 \\ 
  243 & 2002-12-23 & -6.70 & 2002-11-27 & -6.77 \\ 
  244 & 2002-12-28 & -5.43 & 2002-12-08 & -6.44 \\ 
  245 & 2003-01-04 & -4.93 & 2002-12-15 & -6.77 \\ 
  246 & 2003-01-06 & -4.88 & 2002-12-20 & -7.70 \\ 
  247 & 2003-01-11 & -5.50 & 2002-12-23 & -8.75 \\ 
  248 & 2003-01-14 & -4.79 & 2003-01-03 & -4.53 \\ 
  249 & 2003-01-17 & -4.20 & 2003-01-06 & -4.44 \\ 
  250 & 2003-01-24 & -6.63 & 2003-01-11 & -4.85 \\ 
  251 & 2003-01-27 & -7.60 & 2003-01-14 & -4.60 \\ 
  252 & 2003-02-03 & -5.99 & 2003-01-16 & -4.73 \\ 
  253 & 2003-02-13 & -3.94 & 2003-01-18 & -4.73 \\ 
  254 & 2003-02-18 & -5.81 & 2003-01-24 & -7.33 \\ 
  255 & 2003-02-28 & -3.11 & 2003-01-27 & -8.26 \\ 
  256 & 2003-03-02 & -2.95 & 2003-02-03 & -8.06 \\ 
  257 & 2003-03-05 & -3.69 & 2003-02-07 & -4.93 \\ 
  258 & 2003-03-10 & -3.74 & 2003-02-12 & -5.04 \\ 
  259 & 2003-03-15 & -4.11 & 2003-02-18 & -7.47 \\ 
  260 & 2003-03-20 & -5.08 & 2003-02-27 & -4.22 \\ 
  261 & 2003-04-01 & -5.83 & 2003-03-02 & -3.53 \\ 
  262 & 2003-04-06 & -5.14 & 2003-03-07 & -4.64 \\ 
  263 & 2003-04-11 & -8.00 & 2003-03-10 & -5.04 \\ 
  264 & 2003-04-14 & -6.99 & 2003-03-14 & -4.90 \\ 
  265 & 2003-04-23 & -5.45 & 2003-03-20 & -7.80 \\ 
  266 & 2003-04-25 & -5.48 & 2003-03-31 & -7.74 \\ 
  267 & 2003-04-27 & -5.41 & 2003-04-05 & -6.44 \\ 
  268 & 2003-04-30 & -5.77 & 2003-04-10 & -8.45 \\ 
  269 & 2003-05-02 & -5.69 & 2003-04-18 & -5.08 \\ 
  270 & 2003-05-07 & -5.79 & 2003-04-22 & -4.88 \\ 
  271 & 2003-05-09 & -6.14 & 2003-04-25 & -4.86 \\ 
  272 & 2003-05-20 & -4.07 & 2003-05-02 & -6.53 \\ 
  273 & 2003-05-22 & -4.33 & 2003-05-06 & -6.66 \\ 
  274 & 2003-05-31 & -11.37 & 2003-05-09 & -6.47 \\ 
  275 & 2003-06-06 & -6.11 & 2003-05-14 & -5.41 \\ 
  276 & 2003-06-10 & -6.77 & 2003-05-22 & -4.99 \\ 
  277 & 2003-06-24 & -9.38 & 2003-05-30 & -11.93 \\ 
  278 & 2003-06-27 & -7.55 & 2003-06-10 & -7.06 \\ 
  279 & 2003-07-04 & -6.03 & 2003-06-16 & -6.60 \\ 
  280 & 2003-07-07 & -5.41 & 2003-06-23 & -9.88 \\ 
  281 & 2003-07-10 & -5.20 & 2003-06-27 & -8.06 \\ 
  282 & 2003-07-15 & -6.31 & 2003-07-03 & -7.21 \\ 
  283 & 2003-07-20 & -5.07 & 2003-07-07 & -6.08 \\ 
  284 & 2003-07-22 & -5.25 & 2003-07-10 & -6.04 \\ 
  285 & 2003-07-27 & -6.10 & 2003-07-14 & -6.69 \\ 
  286 & 2003-07-30 & -5.99 & 2003-07-17 & -6.51 \\ 
  287 & 2003-08-01 & -6.40 & 2003-07-20 & -5.66 \\
\hline
\end{tabular}
\end{table}

\begin{table}[ht]
%\caption{ \textbf{ FD List from MOSC FD1(\%) and APTY FD2(\%) during 1996-2005}}
\centering
\label*{table 1}
\begin{tabular}{rlrlr}
  \hline
S/N & Date & $FD_{MOSC}$(\%) & Date & $FD_{APTY}$(\%) \\ 
  \hline
 
  288 & 2003-08-05 & -5.83 & 2003-07-23 & -5.66 \\ 
  289 & 2003-08-09 & -5.93 & 2003-07-27 & -6.26 \\ 
  290 & 2003-08-11 & -5.93 & 2003-07-30 & -6.32 \\ 
  291 & 2003-08-18 & -7.07 & 2003-08-05 & -5.30 \\ 
  292 & 2003-08-22 & -5.76 & 2003-08-10 & -5.45 \\ 
  293 & 2003-08-26 & -5.60 & 2003-08-18 & -6.52 \\ 
  294 & 2003-08-30 & -6.08 & 2003-08-21 & -5.42 \\ 
  295 & 2003-09-04 & -4.40 & 2003-08-27 & -5.57 \\ 
  296 & 2003-09-12 & -4.54 & 2003-08-30 & -5.62 \\ 
  297 & 2003-09-18 & -3.97 & 2003-09-04 & -5.34 \\ 
  298 & 2003-09-25 & -2.37 & 2003-09-10 & -5.33 \\ 
  299 & 2003-09-30 & -2.19 & 2003-09-12 & -5.14 \\ 
  300 & 2003-10-03 & -1.27 & 2003-09-14 & -5.14 \\ 
  301 & 2003-10-09 & -3.32 & 2003-09-18 & -5.26 \\ 
  302 & 2003-10-16 & -2.93 & 2003-09-22 & -4.25 \\ 
  303 & 2003-10-23 & -5.55 & 2003-09-27 & -2.42 \\ 
  304 & 2003-10-25 & -7.17 & 2003-09-30 & -2.90 \\ 
  305 & 2003-10-31 & -22.76 & 2003-10-09 & -4.70 \\ 
  306 & 2003-11-07 & -12.30 & 2003-10-15 & -4.04 \\ 
  307 & 2003-11-17 & -10.63 & 2003-10-22 & -6.75 \\ 
  308 & 2003-11-24 & -12.04 & 2003-10-25 & -9.32 \\ 
  309 & 2003-12-01 & -8.54 & 2003-10-31 & -24.75 \\ 
  310 & 2003-12-10 & -7.65 & 2003-11-07 & -13.47 \\ 
  311 & 2003-12-20 & -4.35 & 2003-11-12 & -8.85 \\ 
  312 & 2003-12-23 & -5.11 & 2003-11-17 & -12.83 \\ 
  313 & 2003-12-28 & -4.20 & 2003-11-21 & -11.93 \\ 
  314 & 2004-01-01 & -3.85 & 2003-11-24 & -13.03 \\ 
  315 & 2004-01-04 & -5.84 & 2003-12-04 & -8.01 \\ 
  316 & 2004-01-10 & -9.47 & 2003-12-08 & -7.52 \\ 
  317 & 2004-01-18 & -4.01 & 2003-12-11 & -8.34 \\ 
  318 & 2004-01-25 & -9.23 & 2003-12-14 & -6.63 \\ 
  319 & 2004-01-31 & -6.85 & 2003-12-19 & -5.75 \\ 
  320 & 2004-02-03 & -6.91 & 2003-12-23 & -5.41 \\ 
  321 & 2004-02-13 & -3.66 & 2003-12-25 & -5.86 \\ 
  322 & 2004-03-01 & -3.36 & 2003-12-28 & -5.22 \\ 
  323 & 2004-03-10 & -2.55 & 2004-01-04 & -5.31 \\ 
  324 & 2004-03-13 & -2.66 & 2004-01-10 & -9.60 \\ 
  325 & 2004-03-16 & -2.35 & 2004-01-25 & -8.55 \\ 
  326 & 2004-03-20 & -1.73 & 2004-02-01 & -6.90 \\ 
  327 & 2004-03-22 & -1.64 & 2004-02-04 & -6.85 \\ 
  328 & 2004-03-24 & -1.58 & 2004-02-15 & -3.81 \\ 
  329 & 2004-03-29 & -1.70 & 2004-02-19 & -2.59 \\ 
  330 & 2004-04-01 & -1.71 & 2004-02-22 & -1.09 \\ 
  331 & 2004-04-04 & -3.47 & 2004-03-01 & -2.59 \\ 
  332 & 2004-04-07 & -1.87 & 2004-03-10 & -1.48 \\ 
  333 & 2004-04-11 & -2.24 & 2004-03-12 & -1.49 \\ 
  334 & 2004-04-13 & -1.74 & 2004-03-16 & -1.74 \\
\hline
\end{tabular}
\end{table}

\begin{table}[ht]
%\caption{ \textbf{ FD List from MOSC FD1(\%) and APTY FD2(\%) during 1996-2005}}
\centering
\label*{table 1}
\begin{tabular}{rlrlr}
  \hline
S/N & Date & $FD_{MOSC}$(\%) & Date & $FD_{APTY}$(\%) \\ 
  \hline
 
  335 & 2004-04-15 & -0.92 & 2004-03-21 & -1.56 \\ 
  336 & 2004-04-22 & -1.30 & 2004-03-29 & -0.86 \\ 
  337 & 2004-04-28 & -1.97 & 2004-04-04 & -2.46 \\ 
  338 & 2004-05-09 & -0.60 & 2004-04-11 & -1.93 \\ 
  339 & 2004-05-12 & -0.81 & 2004-04-13 & -0.61 \\ 
  340 & 2004-05-19 & -0.22 & 2004-04-28 & -0.34 \\ 
  341 & 2004-05-25 & -0.11 & 2004-06-10 & -0.20 \\ 
  342 & 2004-05-30 & -0.69 & 2004-06-20 & -0.01 \\ 
  343 & 2004-06-07 & -0.25 & 2004-07-17 & -1.28 \\ 
  344 & 2004-06-10 & -1.01 & 2004-07-20 & -0.28 \\ 
  345 & 2004-06-21 & -0.52 & 2004-07-24 & -3.68 \\ 
  346 & 2004-06-26 & -0.51 & 2004-07-27 & -7.78 \\ 
  347 & 2004-06-30 & -0.03 & 2004-08-01 & -3.67 \\ 
  348 & 2004-07-17 & -0.76 & 2004-08-04 & -3.38 \\ 
  349 & 2004-07-20 & -0.53 & 2004-08-22 & -0.16 \\ 
  350 & 2004-07-24 & -3.80 & 2004-09-15 & -1.75 \\ 
  351 & 2004-07-27 & -8.39 & 2004-09-18 & -1.49 \\ 
  352 & 2004-08-01 & -4.28 & 2004-09-22 & -1.48 \\ 
  353 & 2004-08-04 & -4.65 & 2004-11-10 & -7.23 \\ 
  354 & 2004-08-15 & -0.73 & 2004-11-12 & -6.00 \\ 
  355 & 2004-08-17 & -1.33 & 2004-12-06 & -1.09 \\ 
  356 & 2004-08-22 & -1.14 & 2004-12-08 & -0.48 \\ 
  357 & 2004-08-30 & -0.10 & 2004-12-11 & -0.19 \\ 
  358 & 2004-09-15 & -1.83 & 2004-12-14 & -1.22 \\ 
  359 & 2004-09-18 & -1.85 & 2004-12-30 & -1.04 \\ 
  360 & 2004-09-22 & -0.68 & 2005-01-04 & -4.38 \\ 
  361 & 2004-11-08 & -2.80 & 2005-01-09 & -2.63 \\ 
  362 & 2004-11-10 & -6.26 & 2005-01-19 & -14.44 \\ 
  363 & 2004-12-06 & -1.98 & 2005-01-22 & -10.24 \\ 
  364 & 2004-12-09 & -0.65 & 2005-01-28 & -1.00 \\ 
  365 & 2004-12-13 & -0.82 & 2005-01-31 & -1.20 \\ 
  366 & 2004-12-29 & -1.40 & 2005-02-02 & -1.08 \\ 
  367 & 2005-01-04 & -4.93 & 2005-02-11 & -0.12 \\ 
  368 & 2005-01-09 & -3.49 & 2005-02-19 & -0.61 \\ 
  369 & 2005-01-19 & -13.73 & 2005-05-09 & -3.15 \\ 
  370 & 2005-01-22 & -10.02 & 2005-05-16 & -5.90 \\ 
  371 & 2005-01-28 & -1.44 & 2005-05-30 & -1.22 \\ 
  372 & 2005-01-31 & -1.17 & 2005-06-17 & -0.41 \\ 
  373 & 2005-02-02 & -0.97 & 2005-07-13 & -1.09 \\ 
  374 & 2005-02-04 & -0.67 & 2005-07-17 & -5.57 \\ 
  375 & 2005-02-08 & -0.88 & 2005-08-03 & -0.54 \\ 
  376 & 2005-02-16 & -0.25 & 2005-08-07 & -2.85 \\ 
  377 & 2005-02-19 & -0.73 & 2005-08-10 & -1.01 \\ 
  378 & 2005-02-22 & -0.82 & 2005-08-13 & -0.24 \\ 
  379 & 2005-03-03 & -0.13 & 2005-08-25 & -2.83 \\ 
  380 & 2005-03-06 & -0.51 & 2005-09-03 & -0.91 \\ 
  381 & 2005-03-21 & -0.51 & 2005-09-13 & -10.89 \\
\hline
\end{tabular}
\end{table}

\begin{table}[ht]
%\caption{ \textbf{ FD List from MOSC FD1(\%) and APTY FD2(\%) during 1996-2005}}
\centering
\label*{table 1}
\begin{tabular}{rlrlr}
  \hline
S/N & Date & $FD_{MOSC}$(\%) & Date & $FD_{APTY}$(\%) \\ 
  \hline
 
  382 & 2005-03-25 & -0.49 & 2005-09-15 & -9.84 \\ 
  383 & 2005-03-29 & -0.63 & 2005-09-26 & -0.46 \\ 
  384 & 2005-04-01 & -0.27 &  &  \\ 
  385 & 2005-04-05 & -0.04 &  &  \\ 
  386 & 2005-05-09 & -2.58 &  &  \\ 
  387 & 2005-05-12 & -1.43 &  &  \\ 
  388 & 2005-05-16 & -5.78 &  &  \\ 
  389 & 2005-05-30 & -1.76 &  &  \\ 
  390 & 2005-06-01 & -0.87 &  &  \\ 
  391 & 2005-06-07 & -0.18 &  &  \\ 
  392 & 2005-06-13 & -0.69 &  &  \\ 
  393 & 2005-06-17 & -2.19 &  &  \\ 
  394 & 2005-06-25 & -0.05 &  &  \\ 
  395 & 2005-07-11 & -0.48 &  &  \\ 
  396 & 2005-07-13 & -1.32 &  &  \\ 
  397 & 2005-07-17 & -5.88 &  &  \\ 
  398 & 2005-07-30 & -0.71 &  &  \\ 
  399 & 2005-08-03 & -1.50 &  &  \\ 
  400 & 2005-08-07 & -3.32 &  &  \\ 
  401 & 2005-08-10 & -1.15 &  &  \\ 
  402 & 2005-08-14 & -0.69 &  &  \\ 
  403 & 2005-08-16 & -0.56 &  &  \\ 
  404 & 2005-08-25 & -2.88 &  &  \\ 
  405 & 2005-09-03 & -0.89 &  &  \\ 
  406 & 2005-09-13 & -11.14 &  &  \\ 
  407 & 2005-09-15 & -9.93 &  &  \\ 
  408 & 2005-10-11 & -0.12 &  &  \\ 
   \hline
\end{tabular}
\end{table}

\begin{table}[ht]

\caption{ \textbf{Simultaneous FD events from MOSC  and APTY  and associated parameters during 1996-2005}}
\label{table 2}
\centering
\begin{tabular}{rlrrrrr}
  \hline 
 S/N & Date & $FD_{MOSC}$(\%) & $FD_{APTY}$(\%) & IMF & SWS & Dst \\ 
  \hline
1 & 1998-05-02 & -1.78 & -1.34 & 14.50 & 601 & -36 \\ 
  2 & 1998-08-27 & -3.16 & -2.08 & 14.10 & 630 & -129 \\ 
  3 & 1999-02-18 & -3.46 & -2.96 & 17.10 & 599 & -84 \\ 
  4 & 1999-08-22 & -2.57 & -1.61 & 6.00 & 428 & -27 \\ 
  5 & 1999-08-25 & -1.72 & -0.05 & 7.70 & 538 & -15 \\ 
  6 & 1999-09-16 & -1.76 & -1.05 & 6.20 & 572 & -46 \\ 
  7 & 1999-09-25 & -0.75 & -0.01 & 5.60 & 409 & -16 \\ 
  8 & 1999-09-29 & -1.53 & -1.50 & 6.80 & 539 & -30 \\ 
  9 & 1999-10-12 & -2.20 & -1.57 & 7.30 & 578 & -48 \\ 
  10 & 1999-11-01 & -2.93 & -2.86 & 7.20 & 440 & -15 \\ 
  11 & 1999-11-09 & -2.43 & -2.37 & 6.30 & 615 & -46 \\ 
  12 & 1999-11-18 & -3.72 & -4.27 & 6.00 & 541 & -31 \\ 
  13 & 1999-11-20 & -3.63 & -4.83 & 8.10 & 443 & -16 \\ 
  14 & 1999-12-02 & -3.29 & -3.64 & 10.00 & 344 &  13 \\ 
  15 & 1999-12-13 & -7.47 & -7.49 & 11.40 & 489 & -46 \\ 
  16 & 1999-12-27 & -4.35 & -4.53 & 7.90 & 410 &   2 \\ 
  17 & 2000-01-07 & -1.73 & -4.64 & 4.50 & 522 & -21 \\ 
  18 & 2000-01-13 & -1.68 & -2.94 & 5.10 & 537 & -23 \\ 
  19 & 2000-01-31 & -1.22 & -4.05 & 4.90 & 585 & -12 \\ 
  20 & 2000-02-12 & -5.31 & -6.44 & 14.70 & 553 & -76 \\ 
  21 & 2000-02-21 & -3.90 & -4.67 & 14.30 & 423 &  -1 \\ 
  22 & 2000-03-20 & -3.65 & -4.93 & 7.30 & 348 &   7 \\ 
  23 & 2000-03-24 & -5.18 & -6.37 & 6.60 & 649 &  -3 \\ 
  24 & 2000-03-30 & -4.57 & -5.79 & 5.20 & 446 &  -2 \\ 
  25 & 2000-04-07 & -6.15 & -6.84 & 9.90 & 573 & -162 \\ 
  26 & 2000-04-17 & -3.90 & -4.20 & 6.20 & 457 & -23 \\ 
  27 & 2000-04-20 & -4.70 & -4.60 & 5.90 & 503 & -13 \\ 
  28 & 2000-04-24 & -4.53 & -5.33 & 9.40 & 485 & -25 \\ 
  29 & 2000-05-03 & -4.35 & -6.53 & 6.20 & 520 & -12 \\ 
  30 & 2000-05-08 & -5.12 & -7.04 & 9.80 & 360 &  19 \\ 
  31 & 2000-05-15 & -4.31 & -6.70 & 9.10 & 414 &   7 \\ 
  32 & 2000-05-24 & -8.87 & -10.78 & 13.70 & 636 & -90 \\ 
  33 & 2000-05-30 & -5.66 & -6.84 & 6.20 & 617 & -29 \\ 
  34 & 2000-06-09 & -10.62 & -11.70 & 10.20 & 609 & -34 \\ 
  35 & 2000-06-20 & -6.92 & -8.40 & 6.20 & 379 &   5 \\ 
  36 & 2000-06-24 & -7.21 & -8.78 & 9.00 & 551 & -20 \\ 
  37 & 2000-06-26 & -7.25 & -8.63 & 11.50 & 512 & -36 \\ 
  38 & 2000-07-05 & -4.81 & -6.22 & 5.80 & 449 &   1 \\ 
  39 & 2000-07-11 & -7.44 & -8.63 & 13.70 & 458 &  13 \\ 
  40 & 2000-07-16 & -16.72 & -18.13 & 21.80 & 816 & -172 \\ 
  41 & 2000-07-20 & -12.51 & -13.62 & 8.10 & 533 & -67 \\ 
  42 & 2000-08-06 & -8.90 & -10.59 & 6.00 & 515 & -32 \\ 
  43 & 2000-08-12 & -9.76 & -11.70 & 25.00 & 599 & -128 \\ 
  44 & 2000-08-29 & -5.37 & -7.06 & 6.80 & 596 & -33 \\ 
  45 & 2000-09-03 & -5.64 & -7.04 & 6.80 & 413 & -14 \\ 
  46 & 2000-09-09 & -6.30 & -8.04 & 4.40 & 425 &  -9 \\
\hline
\end{tabular}
\end{table}

\begin{table}[ht]
%\caption{ \textbf{Simultaneous FD List from MOSC FD1(\%) and APTY FD2(\%) during 1996-2005}}
\label*{table 2}
\centering
\begin{tabular}{rlrrrrr}
  \hline 
 S/N & Date & $FD_{MOSC}$(\%) & $FD_{APTY}$(\%) & IMF & SWS & Dst \\ 
  \hline 
 
  47 & 2000-09-18 & -9.62 & -12.03 & 19.20 & 744 & -103 \\ 
  48 & 2000-09-29 & -4.34 & -6.74 & 5.50 & 378 & -19 \\ 
  49 & 2000-10-01 & -4.01 & -6.64 & 4.30 & 418 & -37 \\ 
  50 & 2000-10-05 & -5.01 & -6.62 & 13.40 & 486 & -138 \\ 
  51 & 2000-10-07 & -5.41 & -6.99 & 3.80 & 391 & -36 \\ 
  52 & 2000-10-14 & -4.54 & -6.14 & 12.00 & 411 & -80 \\ 
  53 & 2000-10-29 & -7.10 & -8.11 & 13.70 & 381 & -89 \\ 
  54 & 2000-11-01 & -6.60 & -7.25 & 6.30 & 425 & -16 \\ 
  55 & 2000-11-07 & -8.68 & -9.85 & 20.20 & 512 & -89 \\ 
  56 & 2000-11-11 & -7.02 & -8.44 & 7.20 & 804 & -35 \\ 
  57 & 2000-11-16 & -6.49 & -7.51 & 4.80 & 391 &   2 \\ 
  58 & 2000-11-29 & -10.52 & -12.28 & 9.20 & 512 & -81 \\ 
  59 & 2000-12-03 & -7.54 & -9.43 & 10.00 & 430 &  -9 \\ 
  60 & 2000-12-11 & -5.70 & -7.06 & 4.00 & 552 &   0 \\ 
  61 & 2000-12-25 & -6.06 & -6.97 & 11.90 & 352 &  10 \\ 
  62 & 2000-12-27 & -6.29 & -6.95 & 7.70 & 390 &  -1 \\ 
  63 & 2001-01-03 & -4.85 & -6.34 & 6.80 & 351 &  -8 \\ 
  64 & 2001-01-05 & -4.86 & -6.12 & 6.00 & 403 &  -7 \\ 
  65 & 2001-01-09 & -4.89 & -6.40 & 4.00 & 403 & -13 \\ 
  66 & 2001-03-04 & -2.43 & -3.83 & 6.90 & 448 & -17 \\ 
  67 & 2001-03-20 & -1.43 & -3.00 & 18.00 & 401 & -117 \\ 
  68 & 2001-03-28 & -3.16 & -4.60 & 8.90 & 608 & -53 \\ 
  69 & 2001-04-01 & -5.22 & -6.96 & 7.50 & 746 & -137 \\ 
  70 & 2001-04-05 & -6.53 & -7.04 & 7.50 & 617 & -31 \\ 
  71 & 2001-04-09 & -8.15 & -8.75 & 8.60 & 622 & -53 \\ 
  72 & 2001-04-12 & -14.30 & -14.88 & 15.10 & 659 & -131 \\ 
  73 & 2001-04-16 & -7.08 & -7.78 & 4.00 & 453 & -24 \\ 
  74 & 2001-04-19 & -5.83 & -6.97 & 7.90 & 436 & -41 \\ 
  75 & 2001-04-22 & -4.85 & -5.44 & 11.80 & 360 & -55 \\ 
  76 & 2001-04-29 & -7.99 & -9.52 & 7.60 & 596 & -18 \\ 
  77 & 2001-05-12 & -2.25 & -4.18 & 10.80 & 534 & -35 \\ 
  78 & 2001-05-25 & -4.02 & -5.73 & 6.60 & 557 &   5 \\ 
  79 & 2001-05-28 & -6.17 & -8.11 & 9.10 & 505 &  -8 \\ 
  80 & 2001-06-03 & -3.18 & -4.37 & 5.30 & 508 &  -7 \\ 
  81 & 2001-06-09 & -3.06 & -4.30 & 9.70 & 510 &   0 \\ 
  82 & 2001-06-12 & -3.00 & -4.05 & 5.20 & 433 &  -2 \\ 
  83 & 2001-06-20 & -3.54 & -4.46 & 5.90 & 700 & -20 \\ 
  84 & 2001-06-26 & -2.51 & -3.90 & 6.20 & 464 &  -2 \\ 
  85 & 2001-06-29 & -2.44 & -4.12 & 3.40 & 347 &  11 \\ 
  86 & 2001-07-17 & -2.04 & -2.83 & 8.90 & 592 & -13 \\ 
  87 & 2001-07-30 & -3.12 & -4.23 & 6.80 & 312 &  10 \\ 
  88 & 2001-08-03 & -4.22 & -5.30 & 7.20 & 405 &   0 \\ 
  89 & 2001-08-06 & -4.24 & -5.22 & 7.00 & 440 & -18 \\ 
  90 & 2001-08-10 & -3.00 & -3.94 & 6.60 & 432 &   9 \\ 
  91 & 2001-08-14 & -3.11 & -4.08 & 7.80 & 456 & -10 \\ 
  92 & 2001-08-29 & -8.46 & -9.67 & 4.20 & 459 &  -7 \\ 
  93 & 2001-09-07 & -3.79 & -5.53 & 6.10 & 369 &   6 \\ 
  94 & 2001-09-14 & -3.25 & -4.29 & 10.10 & 414 &   1 \\ 
  95 & 2001-09-19 & -3.01 & -4.48 & 6.50 & 422 &  -5 \\ 
\hline
\end{tabular}
\end{table}

\begin{table}[ht]
%\caption{ \textbf{Simultaneous FD List from MOSC FD1(\%) and APTY FD2(\%) during 1996-2005}}
\label*{table 2}
\centering
\begin{tabular}{rlrrrrr}
  \hline 
 S/N & Date & $FD_{MOSC}$(\%) & $FD_{APTY}$(\%) & IMF & SWS & Dst \\ 
  \hline 

  96 & 2001-09-26 & -8.82 & -10.48 & 10.70 & 549 & -72 \\ 
  97 & 2001-10-02 & -8.58 & -10.14 & 7.50 & 497 & -87 \\ 
  98 & 2001-10-09 & -5.48 & -5.96 & 8.30 & 445 & -37 \\ 
  99 & 2001-10-12 & -6.31 & -7.45 & 11.40 & 501 & -51 \\ 
  100 & 2001-10-22 & -5.61 & -8.44 & 15.10 & 578 & -150 \\ 
  101 & 2001-10-28 & -4.82 & -7.01 & 11.20 & 450 & -99 \\ 
  102 & 2001-11-07 & -6.56 & -9.15 & 6.50 & 635 & -110 \\ 
  103 & 2001-11-25 & -8.61 & -10.76 & 11.50 & 650 & -106 \\ 
  104 & 2001-12-07 & -4.49 & -5.92 & 6.70 & 459 & -19 \\ 
  105 & 2001-12-17 & -3.95 & -5.79 & 8.80 & 471 & -30 \\ 
  106 & 2001-12-22 & -0.94 & -3.49 & 7.70 & 379 & -40 \\ 
  107 & 2001-12-29 & -2.41 & -4.25 & 15.40 & 397 &  34 \\ 
  108 & 2002-01-03 & -7.86 & -9.25 & 5.90 & 342 & -16 \\ 
  109 & 2002-01-11 & -6.33 & -8.44 & 8.90 & 610 & -42 \\ 
  110 & 2002-01-21 & -3.51 & -5.23 & 7.90 & 452 & -10 \\ 
  111 & 2002-02-19 & -0.74 & -3.82 & 7.70 & 401 & -17 \\ 
  112 & 2002-03-05 & -1.83 & -4.01 & 9.50 & 646 & -22 \\ 
  113 & 2002-03-12 & -1.96 & -3.61 & 7.90 & 453 &  -3 \\ 
  114 & 2002-03-30 & -3.43 & -5.94 & 11.60 & 521 &  -5 \\ 
  115 & 2002-04-06 & -1.94 & -4.16 & 6.80 & 358 &   4 \\ 
  116 & 2002-04-12 & -2.92 & -5.15 & 8.70 & 432 &  -1 \\ 
  117 & 2002-04-15 & -1.72 & -4.11 & 8.80 & 357 &  -8 \\ 
  118 & 2002-04-18 & -4.86 & -6.55 & 12.80 & 485 & -104 \\ 
  119 & 2002-04-20 & -5.63 & -7.53 & 10.10 & 563 & -106 \\ 
  120 & 2002-04-24 & -4.95 & -7.33 & 7.10 & 488 & -30 \\ 
  121 & 2002-05-08 & -1.64 & -3.20 & 8.50 & 366 & -19 \\ 
  122 & 2002-05-15 & -3.73 & -5.49 & 6.30 & 411 & -43 \\ 
  123 & 2002-05-23 & -5.83 & -7.06 & 17.00 & 606 & -38 \\ 
  124 & 2002-06-11 & -3.84 & -5.59 & 7.70 & 386 & -19 \\ 
  125 & 2002-06-19 & -3.65 & -5.84 & 10.60 & 468 &  -2 \\ 
  126 & 2002-06-29 & -1.62 & -3.56 & 5.10 & 344 &   6 \\ 
  127 & 2002-07-08 & -3.65 & -5.07 & 6.50 & 391 &  -6 \\ 
  128 & 2002-07-20 & -6.52 & -8.32 & 7.40 & 789 & -20 \\ 
  129 & 2002-07-23 & -5.46 & -7.15 & 4.80 & 472 & -12 \\ 
  130 & 2002-07-30 & -9.02 & -10.43 & 7.50 & 422 &   5 \\ 
  131 & 2002-08-09 & -5.63 & -7.41 & 8.20 & 397 &  -7 \\ 
  132 & 2002-08-20 & -6.95 & -8.66 & 7.20 & 479 & -48 \\ 
  133 & 2002-08-23 & -6.80 & -8.74 & 8.80 & 402 & -18 \\ 
  134 & 2002-08-28 & -7.57 & -9.37 & 8.90 & 447 & -19 \\ 
  135 & 2002-09-08 & -6.06 & -7.73 & 11.70 & 479 & -101 \\ 
  136 & 2002-09-24 & -5.87 & -6.86 & 9.30 & 376 &  -8 \\ 
  137 & 2002-10-01 & -4.86 & -5.93 & 19.50 & 388 & -100 \\ 
  138 & 2002-10-03 & -4.91 & -6.86 & 11.50 & 464 & -78 \\ 
  139 & 2002-10-13 & -2.60 & -3.86 & 6.50 & 301 & -30 \\ 
  140 & 2002-10-25 & -5.27 & -7.10 & 6.80 & 689 & -68 \\ 
  141 & 2002-11-03 & -5.26 & -7.26 & 9.70 & 478 & -65 \\ 
  142 & 2002-11-05 & -6.09 & -8.32 & 8.40 & 545 & -46 \\ 
  143 & 2002-11-12 & -7.79 & -9.43 & 12.40 & 569 & -15 \\ 
  144 & 2002-11-18 & -9.28 & -10.98 & 9.30 & 378 & -37 \\
\hline
\end{tabular}
\end{table}

\begin{table}[ht]
%\caption{ \textbf{Simultaneous FD List from MOSC FD1(\%) and APTY FD2(\%) during 1996-2005}}
\label*{table 2}
\centering
\begin{tabular}{rlrrrrr}
  \hline 
 S/N & Date & $FD_{MOSC}$(\%) & $FD_{APTY}$(\%) & IMF & SWS & Dst \\ 
  \hline 
 
  145 & 2002-11-25 & -4.59 & -5.88 & 7.00 & 460 & -46 \\ 
  146 & 2002-11-27 & -5.47 & -6.77 & 9.80 & 538 & -50 \\ 
  147 & 2002-12-08 & -4.74 & -6.44 & 7.10 & 599 & -28 \\ 
  148 & 2002-12-20 & -5.87 & -7.70 & 6.10 & 528 & -47 \\ 
  149 & 2002-12-23 & -6.70 & -8.75 & 10.10 & 517 & -42 \\ 
  150 & 2003-01-06 & -4.88 & -4.44 & 6.10 & 395 &  -2 \\ 
  151 & 2003-01-11 & -5.50 & -4.85 & 7.70 & 435 & -18 \\ 
  152 & 2003-01-14 & -4.79 & -4.60 & 9.40 & 384 & -10 \\ 
  153 & 2003-01-24 & -6.63 & -7.33 & 7.00 & 686 & -20 \\ 
  154 & 2003-01-27 & -7.60 & -8.26 & 8.70 & 507 &  -5 \\ 
  155 & 2003-02-03 & -5.99 & -8.06 & 8.90 & 484 & -42 \\ 
  156 & 2003-02-18 & -5.81 & -7.47 & 8.50 & 652 &  -1 \\ 
  157 & 2003-03-02 & -2.95 & -3.53 & 5.90 & 404 & -24 \\ 
  158 & 2003-03-10 & -3.74 & -5.04 & 6.70 & 400 & -23 \\ 
  159 & 2003-03-20 & -5.08 & -7.80 & 10.60 & 694 & -29 \\ 
  160 & 2003-04-25 & -5.48 & -4.86 & 6.40 & 543 & -42 \\ 
  161 & 2003-05-02 & -5.69 & -6.53 & 4.70 & 583 & -24 \\ 
  162 & 2003-05-09 & -6.14 & -6.47 & 8.30 & 791 & -27 \\ 
  163 & 2003-05-22 & -4.33 & -4.99 & 7.00 & 493 & -42 \\ 
  164 & 2003-06-10 & -6.77 & -7.06 & 6.30 & 695 & -21 \\ 
  165 & 2003-06-27 & -7.55 & -8.06 & 7.50 & 692 & -17 \\ 
  166 & 2003-07-07 & -5.41 & -6.08 & 5.60 & 569 & -16 \\ 
  167 & 2003-07-10 & -5.20 & -6.04 & 6.20 & 356 &  12 \\ 
  168 & 2003-07-20 & -5.07 & -5.66 & 6.00 & 632 & -27 \\ 
  169 & 2003-07-27 & -6.10 & -6.26 & 8.80 & 677 & -36 \\ 
  170 & 2003-07-30 & -5.99 & -6.32 & 6.90 & 763 & -27 \\ 
  171 & 2003-08-05 & -5.83 & -5.30 & 9.00 & 444 &   7 \\ 
  172 & 2003-08-18 & -7.07 & -6.52 & 18.50 & 468 & -108 \\ 
  173 & 2003-08-30 & -6.08 & -5.62 & 5.20 & 544 & -15 \\ 
  174 & 2003-09-04 & -4.40 & -5.34 & 8.50 & 612 & -13 \\ 
  175 & 2003-09-12 & -4.54 & -5.14 & 4.70 & 593 &  -6 \\ 
  176 & 2003-09-18 & -3.97 & -5.26 & 6.30 & 766 & -41 \\ 
  177 & 2003-10-09 & -3.32 & -4.70 & 5.40 & 560 &  -2 \\ 
  178 & 2003-10-25 & -7.17 & -9.32 & 15.40 & 540 & -26 \\ 
  179 & 2003-10-31 & -22.76 & -24.75 & 15.80 & 1003 & -117 \\ 
  180 & 2003-11-07 & -12.30 & -13.47 & 5.80 & 509 &  -9 \\ 
  181 & 2003-11-17 & -10.63 & -12.83 & 6.00 & 750 & -35 \\ 
  182 & 2003-11-24 & -12.04 & -13.03 & 9.10 & 550 & -29 \\ 
  183 & 2003-12-23 & -5.11 & -5.41 & 4.40 & 526 &  -7 \\ 
  184 & 2003-12-28 & -4.20 & -5.22 & 9.70 & 508 & -14 \\ 
  185 & 2004-01-04 & -5.84 & -5.31 & 8.60 & 570 & -21 \\ 
  186 & 2004-01-10 & -9.47 & -9.60 & 11.30 & 551 & -24 \\ 
  187 & 2004-01-25 & -9.23 & -8.55 & 9.90 & 472 & -65 \\ 
  188 & 2004-03-01 & -3.36 & -2.59 & 6.10 & 649 & -14 \\ 
  189 & 2004-03-10 & -2.55 & -1.48 & 8.40 & 694 & -52 \\ 
  190 & 2004-03-16 & -2.35 & -1.74 & 5.40 & 452 & -18 \\ 
  191 & 2004-03-29 & -1.70 & -0.86 & 5.00 & 612 & -11 \\ 
\hline
\end{tabular}
\end{table}

\begin{table}[ht]
%\caption{ \textbf{Simultaneous FD List from MOSC FD1(\%) and APTY FD2(\%) during 1996-2005}}
\label*{table 2}
\centering
\begin{tabular}{rlrrrrr}
  \hline 
 S/N & Date & $FD_{MOSC}$(\%) & $FD_{APTY}$(\%) & IMF & SWS & Dst \\ 
  \hline 

  192 & 2004-04-04 & -3.47 & -2.46 & 14.60 & 456 & -40 \\ 
  193 & 2004-04-11 & -2.24 & -1.93 & 5.00 & 432 & -14 \\ 
  194 & 2004-04-13 & -1.74 & -0.61 & 3.50 & 468 & -10 \\ 
  195 & 2004-04-28 & -1.97 & -0.34 & 7.70 & 481 &   6 \\ 
  196 & 2004-06-10 & -1.01 & -0.20 & 6.40 & 477 &   2 \\ 
  197 & 2004-07-17 & -0.76 & -1.28 & 7.10 & 505 & -39 \\ 
  198 & 2004-07-20 & -0.53 & -0.28 & 6.00 & 527 &  -4 \\ 
  199 & 2004-07-24 & -3.80 & -3.68 & 16.90 & 561 & -13 \\ 
  200 & 2004-07-27 & -8.39 & -7.78 & 17.40 & 904 & -120 \\ 
  201 & 2004-08-01 & -4.28 & -3.67 & 6.50 & 471 & -25 \\ 
  202 & 2004-08-04 & -4.65 & -3.38 & 5.60 & 334 & -10 \\ 
  203 & 2004-08-22 & -1.14 & -0.16 & 4.70 & 450 & -22 \\ 
  204 & 2004-09-15 & -1.83 & -1.75 & 4.90 & 549 & -23 \\ 
  205 & 2004-09-18 & -1.85 & -1.49 & 6.00 & 441 & -17 \\ 
  206 & 2004-09-22 & -0.68 & -1.48 & 7.20 & 477 & -11 \\ 
  207 & 2004-11-10 & -6.26 & -7.23 & 18.40 & 691 & -176 \\ 
  208 & 2004-12-06 & -1.98 & -1.09 & 9.70 & 424 & -26 \\ 
  209 & 2005-01-04 & -4.93 & -4.38 & 5.70 & 713 & -25 \\ 
  210 & 2005-01-09 & -3.49 & -2.63 & 8.60 & 460 & -23 \\ 
  211 & 2005-01-19 & -13.73 & -14.44 & 12.60 & 840 & -64 \\ 
  212 & 2005-01-22 & -10.02 & -10.24 & 13.20 & 766 & -72 \\ 
  213 & 2005-01-28 & -1.44 & -1.00 & 7.40 & 379 &  -3 \\ 
  214 & 2005-01-31 & -1.17 & -1.20 & 7.80 & 611 & -17 \\ 
  215 & 2005-02-02 & -0.97 & -1.08 & 6.30 & 507 & -11 \\ 
  216 & 2005-02-19 & -0.73 & -0.61 & 6.40 & 497 & -25 \\ 
  217 & 2005-05-09 & -2.58 & -3.15 & 8.40 & 620 & -48 \\ 
  218 & 2005-05-16 & -5.78 & -5.90 & 10.20 & 638 & -85 \\ 
  219 & 2005-05-30 & -1.76 & -1.22 & 15.70 & 469 & -73 \\ 
  220 & 2005-06-17 & -2.19 & -0.41 & 7.20 & 573 & -27 \\ 
  221 & 2005-07-13 & -1.32 & -1.09 & 6.90 & 560 & -32 \\ 
  222 & 2005-07-17 & -5.88 & -5.57 & 10.00 & 457 &  -9 \\ 
  223 & 2005-08-03 & -1.50 & -0.54 & 5.00 & 449 &  -9 \\ 
  224 & 2005-08-07 & -3.32 & -2.85 & 5.20 & 657 & -25 \\ 
  225 & 2005-08-10 & -1.15 & -1.01 & 5.50 & 433 & -26 \\ 
  226 & 2005-08-25 & -2.88 & -2.83 & 5.30 & 664 & -71 \\ 
  227 & 2005-09-03 & -0.89 & -0.91 & 6.90 & 596 & -51 \\ 
  228 & 2005-09-13 & -11.14 & -10.89 & 6.00 & 722 & -76 \\ 
  229 & 2005-09-15 & -9.93 & -9.84 & 7.80 & 684 & -49 \\ 
   \hline
\end{tabular}
\end{table}

\begin{table}[ht]

\caption{ \textbf{Simultaneous Large $FDs_{MOSC}$(\%) $\leq-4$ Events and associated solar-geophysical parameters during 1996-2005}}
\label{table 3}
\centering
\begin{tabular}{rlrrrr}
  \hline 
 S/N & Date & $FD_{MOSC}$(\%) & IMF & SWS & Dst \\ 
  \hline

1 & 2000-10-01 & -4.01 & 4.30 & 418 & -37 \\ 
  2 & 2001-05-25 & -4.02 & 6.60 & 557 &   5 \\ 
  3 & 2003-12-28 & -4.20 & 9.70 & 508 & -14 \\ 
  4 & 2001-08-03 & -4.22 & 7.20 & 405 &   0 \\ 
  5 & 2001-08-06 & -4.24 & 7.00 & 440 & -18 \\ 
  6 & 2004-08-01 & -4.28 & 6.50 & 471 & -25 \\ 
  7 & 2000-05-15 & -4.31 & 9.10 & 414 &   7 \\ 
  8 & 2003-05-22 & -4.33 & 7.00 & 493 & -42 \\ 
  9 & 2000-09-29 & -4.34 & 5.50 & 378 & -19 \\ 
  10 & 2000-05-03 & -4.35 & 6.20 & 520 & -12 \\ 
  11 & 1999-12-27 & -4.35 & 7.90 & 410 &   2 \\ 
  12 & 2003-09-04 & -4.40 & 8.50 & 612 & -13 \\ 
  13 & 2001-12-07 & -4.49 & 6.70 & 459 & -19 \\ 
  14 & 2000-04-24 & -4.53 & 9.40 & 485 & -25 \\ 
  15 & 2003-09-12 & -4.54 & 4.70 & 593 &  -6 \\ 
  16 & 2000-10-14 & -4.54 & 12.00 & 411 & -80 \\ 
  17 & 2000-03-30 & -4.57 & 5.20 & 446 &  -2 \\ 
  18 & 2002-11-25 & -4.59 & 7.00 & 460 & -46 \\ 
  19 & 2004-08-04 & -4.65 & 5.60 & 334 & -10 \\ 
  20 & 2000-04-20 & -4.70 & 5.90 & 503 & -13 \\ 
  21 & 2002-12-08 & -4.74 & 7.10 & 599 & -28 \\ 
  22 & 2003-01-14 & -4.79 & 9.40 & 384 & -10 \\ 
  23 & 2000-07-05 & -4.81 & 5.80 & 449 &   1 \\ 
  24 & 2001-10-28 & -4.82 & 11.20 & 450 & -99 \\ 
  25 & 2001-04-22 & -4.85 & 11.80 & 360 & -55 \\ 
  26 & 2001-01-03 & -4.85 & 6.80 & 351 &  -8 \\ 
  27 & 2002-10-01 & -4.86 & 19.50 & 388 & -100 \\ 
  28 & 2002-04-18 & -4.86 & 12.80 & 485 & -104 \\ 
  29 & 2001-01-05 & -4.86 & 6.00 & 403 &  -7 \\ 
  30 & 2003-01-06 & -4.88 & 6.10 & 395 &  -2 \\ 
  31 & 2001-01-09 & -4.89 & 4.00 & 403 & -13 \\ 
  32 & 2002-10-03 & -4.91 & 11.50 & 464 & -78 \\ 
  33 & 2005-01-04 & -4.93 & 5.70 & 713 & -25 \\ 
  34 & 2002-04-24 & -4.95 & 7.10 & 488 & -30 \\ 
  35 & 2000-10-05 & -5.01 & 13.40 & 486 & -138 \\ 
  36 & 2003-07-20 & -5.07 & 6.00 & 632 & -27 \\ 
  37 & 2003-03-20 & -5.08 & 10.60 & 694 & -29 \\ 
  38 & 2003-12-23 & -5.11 & 4.40 & 526 &  -7 \\ 
  39 & 2000-05-08 & -5.12 & 9.80 & 360 &  19 \\ 
  40 & 2000-03-24 & -5.18 & 6.60 & 649 &  -3 \\ 
  41 & 2003-07-10 & -5.20 & 6.20 & 356 &  12 \\ 
  42 & 2001-04-01 & -5.22 & 7.50 & 746 & -137 \\ 
  43 & 2002-11-03 & -5.26 & 9.70 & 478 & -65 \\ 
  44 & 2002-10-25 & -5.27 & 6.80 & 689 & -68 \\
\hline
\end{tabular}
\end{table}
\begin{table}[ht]
%\caption{ \textbf{Simultaneous FD List from MOSC FD1(\%) during 1996-2005}}
\label*{table 3}
\centering
\begin{tabular}{rlrrrr}
  \hline 
 S/N & Date & $FD_{MOSC}$(\%) & IMF & SWS & Dst \\ 
  \hline
 
  45 & 2000-02-12 & -5.31 & 14.70 & 553 & -76 \\ 
  46 & 2000-08-29 & -5.37 & 6.80 & 596 & -33 \\ 
  47 & 2003-07-07 & -5.41 & 5.60 & 569 & -16 \\ 
  48 & 2000-10-07 & -5.41 & 3.80 & 391 & -36 \\ 
  49 & 2002-07-23 & -5.46 & 4.80 & 472 & -12 \\ 
  50 & 2002-11-27 & -5.47 & 9.80 & 538 & -50 \\ 
  51 & 2003-04-25 & -5.48 & 6.40 & 543 & -42 \\ 
  52 & 2001-10-09 & -5.48 & 8.30 & 445 & -37 \\ 
  53 & 2003-01-11 & -5.50 & 7.70 & 435 & -18 \\ 
  54 & 2001-10-22 & -5.61 & 15.10 & 578 & -150 \\ 
  55 & 2002-08-09 & -5.63 & 8.20 & 397 &  -7 \\ 
  56 & 2002-04-20 & -5.63 & 10.10 & 563 & -106 \\ 
  57 & 2000-09-03 & -5.64 & 6.80 & 413 & -14 \\ 
  58 & 2000-05-30 & -5.66 & 6.20 & 617 & -29 \\ 
  59 & 2003-05-02 & -5.69 & 4.70 & 583 & -24 \\ 
  60 & 2000-12-11 & -5.70 & 4.00 & 552 &   0 \\ 
  61 & 2005-05-16 & -5.78 & 10.20 & 638 & -85 \\ 
  62 & 2003-02-18 & -5.81 & 8.50 & 652 &  -1 \\ 
  63 & 2003-08-05 & -5.83 & 9.00 & 444 &   7 \\ 
  64 & 2002-05-23 & -5.83 & 17.00 & 606 & -38 \\ 
  65 & 2001-04-19 & -5.83 & 7.90 & 436 & -41 \\ 
  66 & 2004-01-04 & -5.84 & 8.60 & 570 & -21 \\ 
  67 & 2002-12-20 & -5.87 & 6.10 & 528 & -47 \\ 
  68 & 2002-09-24 & -5.87 & 9.30 & 376 &  -8 \\ 
  69 & 2005-07-17 & -5.88 & 10.00 & 457 &  -9 \\ 
  70 & 2003-07-30 & -5.99 & 6.90 & 763 & -27 \\ 
  71 & 2003-02-03 & -5.99 & 8.90 & 484 & -42 \\ 
  72 & 2002-09-08 & -6.06 & 11.70 & 479 & -101 \\ 
  73 & 2000-12-25 & -6.06 & 11.90 & 352 &  10 \\ 
  74 & 2003-08-30 & -6.08 & 5.20 & 544 & -15 \\ 
  75 & 2002-11-05 & -6.09 & 8.40 & 545 & -46 \\ 
  76 & 2003-07-27 & -6.10 & 8.80 & 677 & -36 \\ 
  77 & 2003-05-09 & -6.14 & 8.30 & 791 & -27 \\ 
  78 & 2000-04-07 & -6.15 & 9.90 & 573 & -162 \\ 
  79 & 2001-05-28 & -6.17 & 9.10 & 505 &  -8 \\ 
  80 & 2004-11-10 & -6.26 & 18.40 & 691 & -176 \\ 
  81 & 2000-12-27 & -6.29 & 7.70 & 390 &  -1 \\ 
  82 & 2000-09-09 & -6.30 & 4.40 & 425 &  -9 \\ 
  83 & 2001-10-12 & -6.31 & 11.40 & 501 & -51 \\ 
  84 & 2002-01-11 & -6.33 & 8.90 & 610 & -42 \\ 
  85 & 2000-11-16 & -6.49 & 4.80 & 391 &   2 \\ 
  86 & 2002-07-20 & -6.52 & 7.40 & 789 & -20 \\ 
  87 & 2001-04-05 & -6.53 & 7.50 & 617 & -31 \\ 
  88 & 2001-11-07 & -6.56 & 6.50 & 635 & -110 \\ 
  89 & 2000-11-01 & -6.60 & 6.30 & 425 & -16 \\ 
  90 & 2003-01-24 & -6.63 & 7.00 & 686 & -20 \\ 
  91 & 2002-12-23 & -6.70 & 10.10 & 517 & -42 \\ 
  92 & 2003-06-10 & -6.77 & 6.30 & 695 & -21 \\ 
  93 & 2002-08-23 & -6.80 & 8.80 & 402 & -18 \\
\hline
\end{tabular}
\end{table}
\begin{table}[ht]
%\caption{ \textbf{Simultaneous FD List from MOSC FD1(\%) during 1996-2005}}
\label*{table 3}
\centering
\begin{tabular}{rlrrrr}
  \hline 
 S/N & Date & $FD_{MOSC}$(\%) & IMF & SWS & Dst \\ 
  \hline
 
  94 & 2000-06-20 & -6.92 & 6.20 & 379 &   5 \\ 
  95 & 2002-08-20 & -6.95 & 7.20 & 479 & -48 \\ 
  96 & 2000-11-11 & -7.02 & 7.20 & 804 & -35 \\ 
  97 & 2003-08-18 & -7.07 & 18.50 & 468 & -108 \\ 
  98 & 2001-04-16 & -7.08 & 4.00 & 453 & -24 \\ 
  99 & 2000-10-29 & -7.10 & 13.70 & 381 & -89 \\ 
  100 & 2003-10-25 & -7.17 & 15.40 & 540 & -26 \\ 
  101 & 2000-06-24 & -7.21 & 9.00 & 551 & -20 \\ 
  102 & 2000-06-26 & -7.25 & 11.50 & 512 & -36 \\ 
  103 & 2000-07-11 & -7.44 & 13.70 & 458 &  13 \\ 
  104 & 1999-12-13 & -7.47 & 11.40 & 489 & -46 \\ 
  105 & 2000-12-03 & -7.54 & 10.00 & 430 &  -9 \\ 
  106 & 2003-06-27 & -7.55 & 7.50 & 692 & -17 \\ 
  107 & 2002-08-28 & -7.57 & 8.90 & 447 & -19 \\ 
  108 & 2003-01-27 & -7.60 & 8.70 & 507 &  -5 \\ 
  109 & 2002-11-12 & -7.79 & 12.40 & 569 & -15 \\ 
  110 & 2002-01-03 & -7.86 & 5.90 & 342 & -16 \\ 
  111 & 2001-04-29 & -7.99 & 7.60 & 596 & -18 \\ 
  112 & 2001-04-09 & -8.15 & 8.60 & 622 & -53 \\ 
  113 & 2004-07-27 & -8.39 & 17.40 & 904 & -120 \\ 
  114 & 2001-08-29 & -8.46 & 4.20 & 459 &  -7 \\ 
  115 & 2001-10-02 & -8.58 & 7.50 & 497 & -87 \\ 
  116 & 2001-11-25 & -8.61 & 11.50 & 650 & -106 \\ 
  117 & 2000-11-07 & -8.68 & 20.20 & 512 & -89 \\ 
  118 & 2001-09-26 & -8.82 & 10.70 & 549 & -72 \\ 
  119 & 2000-05-24 & -8.87 & 13.70 & 636 & -90 \\ 
  120 & 2000-08-06 & -8.90 & 6.00 & 515 & -32 \\ 
  121 & 2002-07-30 & -9.02 & 7.50 & 422 &   5 \\ 
  122 & 2004-01-25 & -9.23 & 9.90 & 472 & -65 \\ 
  123 & 2002-11-18 & -9.28 & 9.30 & 378 & -37 \\ 
  124 & 2004-01-10 & -9.47 & 11.30 & 551 & -24 \\ 
  125 & 2000-09-18 & -9.62 & 19.20 & 744 & -103 \\ 
  126 & 2000-08-12 & -9.76 & 25.00 & 599 & -128 \\ 
  127 & 2005-09-15 & -9.93 & 7.80 & 684 & -49 \\ 
  128 & 2005-01-22 & -10.02 & 13.20 & 766 & -72 \\ 
  129 & 2000-11-29 & -10.52 & 9.20 & 512 & -81 \\ 
  130 & 2000-06-09 & -10.62 & 10.20 & 609 & -34 \\ 
  131 & 2003-11-17 & -10.63 & 6.00 & 750 & -35 \\ 
  132 & 2005-09-13 & -11.14 & 6.00 & 722 & -76 \\ 
  133 & 2003-11-24 & -12.04 & 9.10 & 550 & -29 \\ 
  134 & 2003-11-07 & -12.30 & 5.80 & 509 &  -9 \\ 
  135 & 2000-07-20 & -12.51 & 8.10 & 533 & -67 \\ 
  136 & 2005-01-19 & -13.73 & 12.60 & 840 & -64 \\ 
  137 & 2001-04-12 & -14.30 & 15.10 & 659 & -131 \\ 
  138 & 2000-07-16 & -16.72 & 21.80 & 816 & -172 \\ 
  139 & 2003-10-31 & -22.76 & 15.80 & 1003 & -117 \\ 
   \hline
\end{tabular}
\end{table}

\begin{figure}[h!]
 \centering
  \includegraphics[width=1.36\textwidth]{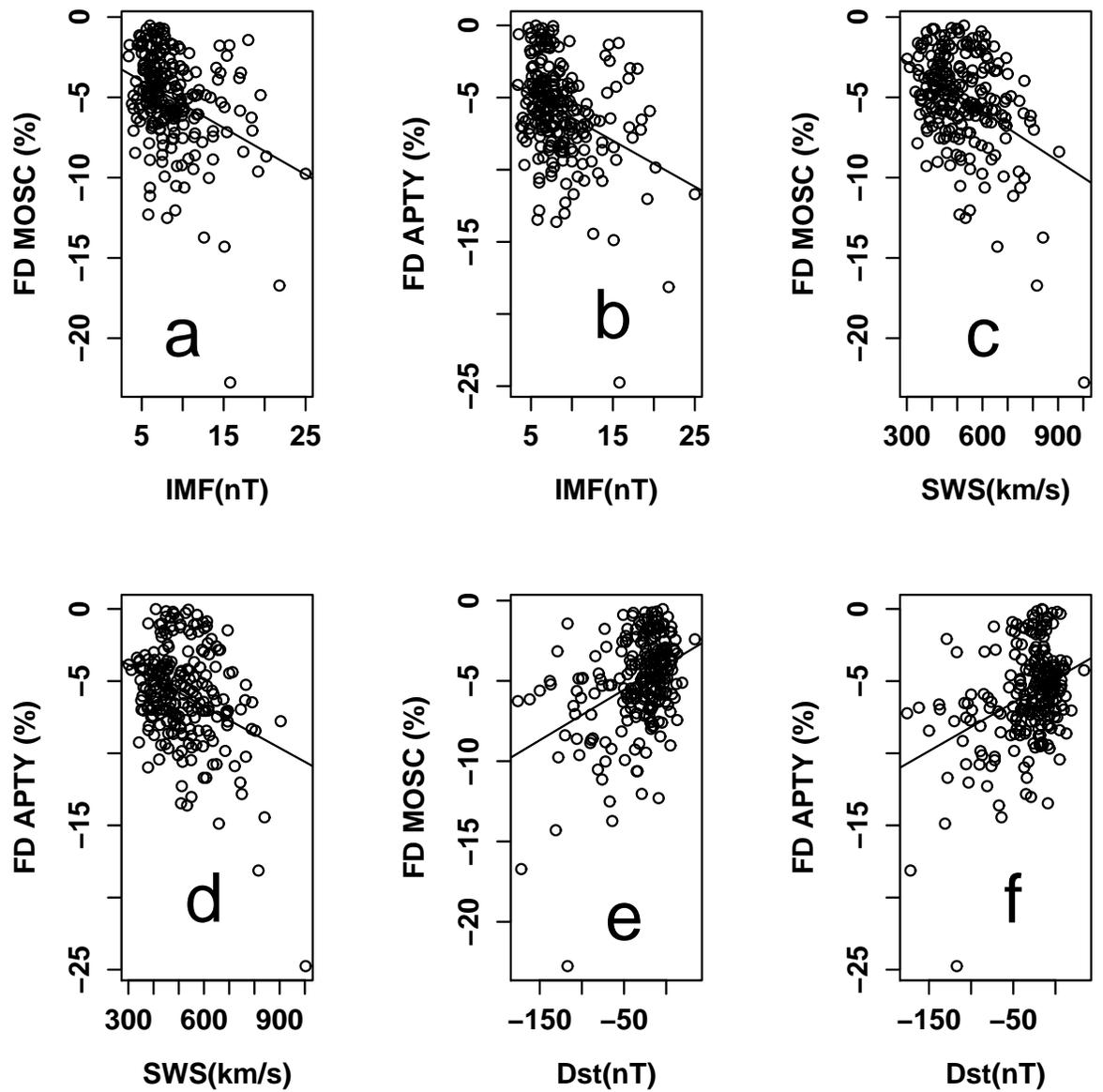}
 \caption{\textbf{Plots of MOSC, APTY FDs and Related Solar Wind Parameters and Geomagnetic Activity Index}}
 \end{figure}

\begin{figure}[h!]
 \centering
  \includegraphics[width=1.36\textwidth]{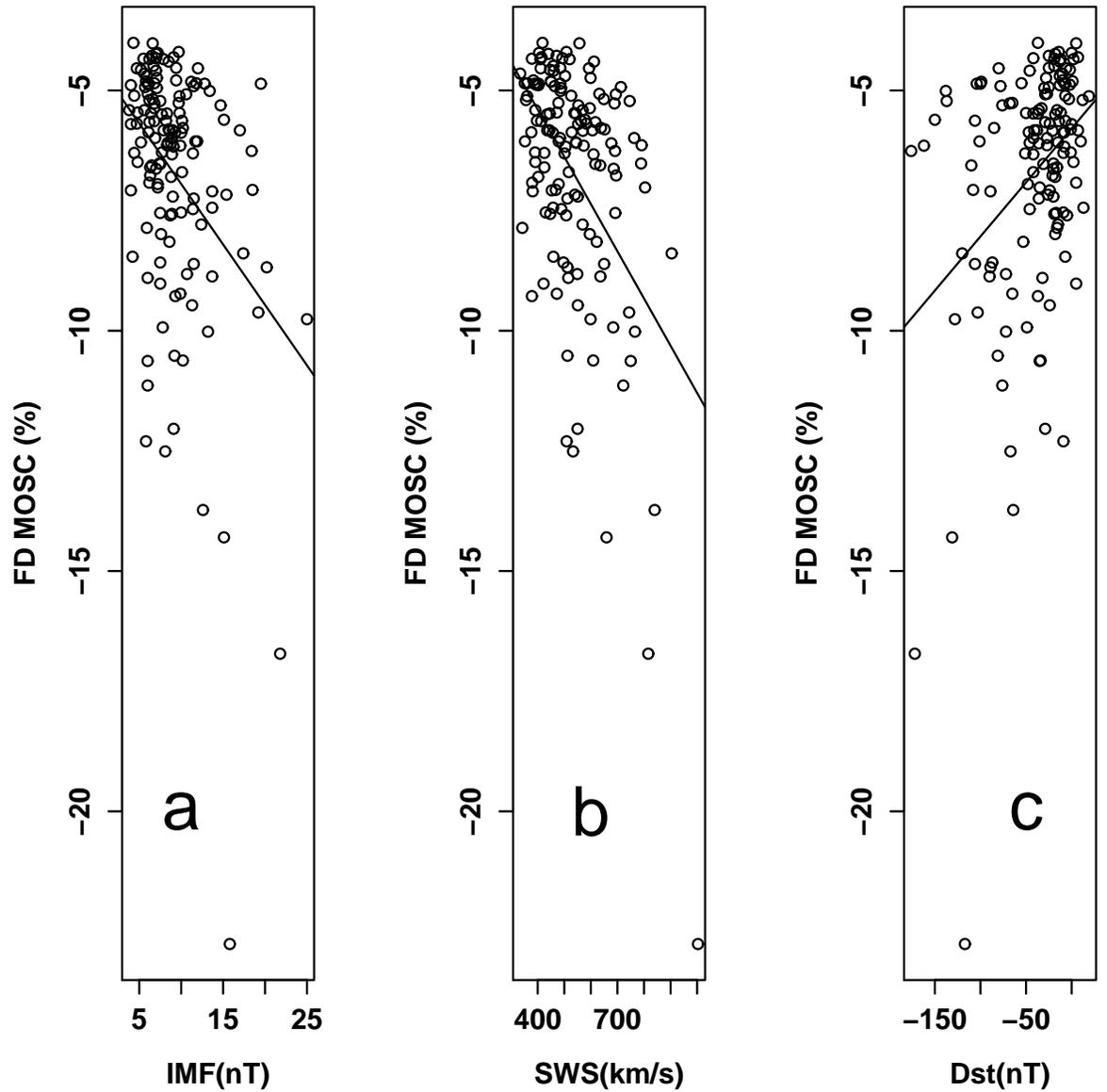}
 \caption{\textbf{Plots of Large $FDs_{MOSC}$(\%) $\leq-4$ and Related Solar Wind Parameters and Geomagnetic Storm Index}}
 \end{figure}

\begin{table}[ht]
    \begin{minipage}[b]{0.485\linewidth}
\caption{ Correlation results of all simultaneous $FD_{MOSC}$, $FD_{APTY}$ and associated characteristics}
\label{table 4}
\centering
\begin{tabular}{rrlrrr}
  \hline
 S/N & Parameters & $R^2$ &  r & p-values  \\ 
  \hline
\hline

   1 & $FD_{MOSC}$-IMF & 0.13 & 0.36 & $3.16\times 10^{-08}$ \\ 
     2 & $FD_{MOSC}$-SWS & 0.17 & 0.41 & $1.27\times 10^{-10}$  \\ 
     3 & $FD_{MOSC}$-Dst & 0.16 & 0.40 & $6.51\times 10^{-10}$  \\ 
    4 & $FD_{APTY}$-IMF & 0.12 & 0.35 & $8.53\times 10^{-08}$  \\ 
     5 & $FD_{APTY}$-SWS & 0.11 & 0.34 & $1.45\times 10^{-07}$  \\ 
     6 & $FD_{APTY}$-Dst & 0.14 & 0.37 & $6.20\times 10^{-09}$  \\ 
\hline
 \end{tabular}
    \end{minipage}
    \hspace{2.4cm}
    \begin{minipage}[b]{0.485\linewidth}
\caption{ Correlation results of large $FDs_{MOSC}$(\%) $\leq-4$ and associated parameters}
\label{table 5}
\centering
\begin{tabular}{rlrrr}
  \hline
 S/N & Parameters & $R^2$ &  r & p-values \\ 
  \hline
\hline
1 & $FD_{MOSC}$-IMF & 0.14 & 0.37 & $5.37\times 10^{-06}$ \\ 
  2 & $FD_{MOSC}$-SWS & 0.23 & 0.48 & $2.13\times 10^{-09}$ \\ 
  3 & $FD_{MOSC}$-Dst & 0.13 & 0.36 & $1.34\times 10^{-05}$  \\ 
   \hline
 \hline
\end{tabular}
    \end{minipage}
\end{table}

%\section{Conclusion}An   FD location code has been developed and deployed to select FD list using daily averaged  CR raw data  from MOSC and APTY  NM stations during solar cycle 23. A large catalogue of FD as well as very small (-0.01) FD magnitude have been selected by the present program. Our software thus, has shown that  the shortcomings that characterizes the manual method of FD selection can be addressed. The regression analysis carried out on the two datasets showed on average, comparable solar-geophysical impact on cosmic ray intensity modulation. In general, the high statistical significance of the results obtained in this submission, demonstrates that IMF intensity, SWS, geomagnetic storm time index  have significant impact on galactic cosmic ray intensity variation and does not appear to depend on FD amplitude. 

\section {Acknowledgments}

We remain indebted to the team that hosts the websites {http://cr0.izmiran.rssi.ru/ and https://omniweb.gsfc.nasa.gov/html/ow data.html} repository from where we freely downloaded the data for this work. In a very special way, we want to acknowledge   the non-commercial R software  developers. The prompt and enriching  input of the anonymous referee is hereby acknowledged. His comments have had significant effect on the manuscript. Our friends on R4DS learning community, especially Shamsuddeen who first introduced JAA to the platform are heartily acknowledged for their invaluable assisitance.

%\section*{References}

%\bibliographystyle{model5-names}\biboptions{authoryear}
%\bibliographystyle{model1a-num-names.bst}
\bibliography{reference}

\end{document}